\documentclass[seceq]{ptptex}

\usepackage{amsmath}
\usepackage{amsfonts}
\usepackage{latexsym}
\usepackage{amssymb} 
\usepackage{verbatim}
\usepackage{amsthm}
\usepackage{graphicx}
\usepackage{wrapft}

\newcommand{\MR}{{\mathbb R}}

\title{
  Second-order Gauge Invariant Perturbation Theory 
}

\subtitle{
  Perturbative curvatures in the two-parameter case
}

\author{
  Kouji \textsc{Nakamura}%
}

\inst{
  Department of Astronomical Science,
  the Graduate University for Advanced Studies,
  Mitaka, Tokyo 181-8588, Japan.
}

\recdate{October 6, 2004}

\abst{
  Based on the gauge invariant variables proposed in our previous paper
  [K.~Nakamura, Prog.~Theor.~Phys.\ {\bf 110} (2003), 723], some
  formulae for the perturbative curvatures of each order are derived.
  We follow the general framework of the second-order gauge invariant
  perturbation theory on an arbitrary background spacetime to derive
  these formulae.
  It is found that these perturbative curvatures have the same
  form as those given in the definitions of gauge invariant
  variables for arbitrary perturbatives fields, which were
  proposed in the above paper. 
  As a result, we explicitly see that any perturbative Einstein
  equation can be given in terms of a gauge invariant form.
  We briefly discuss physical situations to which this framework
  should be applied.
}

\begin{document}

\maketitle

\section{Introduction}
\label{sec:intro}


In many theories of physics, realistic situations are often
difficult to describe by an exact solution of a theory,
because theories in physics and their exact solutions are often
too idealized to properly represent natural phenomena.
Given this situation, we have to consider perturbative approaches to
investigate realistic situations.
General relativity is one theory in which the construction of
exact solutions is not so easy.
Though there are many exact solutions to the Einstein
equation\cite{Exact-solutions}, these are often too idealized.
For this reason, general relativistic perturbation theory is a
useful technique to investigate natural 
phenomena\cite{Cosmological-Perturbations,Gerlach_Sengupta}.


In addition to this technical problem, general relativity is
based on the concept of general covariance.
Intuitively, the principle of general covariance states that
there is no preferred coordinate system in nature, though the
notion of general covariance is mathematically included in the
definition of a spacetime manifold in a trivial way.
This is based on the philosophy that coordinate systems are
originally chosen by us, and natural phenomena have nothing to
do with our coordinate system.
Due to this general covariance, the {\it gauge degree of freedom},
which is an unphysical degree of freedom of perturbations,
arises in general relativistic perturbations.
To obtain physically meaningful results, we have to fix this gauge
degrees of freedom or to extract the {\it gauge invariant part of
perturbations}.
A similar situation has been found in recent investigations of the
oscillatory behavior of a gravitating Nambu-Goto
membrane\cite{kouchan-papers,kouchan-flat}, which concern the
dynamical degrees of freedom of extended gravitating objects.


On the other hand, higher-order multi-parameter perturbations, in
which there are two or more small parameters, can be applied to
many physical situations.
One well-known application of two-parameter perturbation theory is
perturbations of a spherical star\cite{Kojima}, in which we choose
the gravitational field of the spherical star as the background
spacetime for the perturbations, one of the parameters for the
perturbations corresponds to the rotation of the star, and the
other is its pulsation amplitude.
The effects due to the rotation-pulsation coupling are described
at higher orders.
A similar perturbation theory on the Minkowski background
spacetime was developed by the present author to study the
comparison of the oscillatory behavior of a gravitating string
with that of a test string\cite{kouchan-flat}.
Even in the one-parameter case, it is interesting to consider
higher-order perturbations.
In particular, Gleiser et al.\cite{Gleiser-Nicasio} reported
that second-order perturbations yield accurate wave forms of
gravitational waves.
Hence, it is worthwhile to investigate higher-order
multi-parameter perturbation theory from a general point of
view.


Motivated by these physical applications, the general relativistic
gauge invariant multi-parameter perturbation theory has been
developed in a number of
papers\cite{kouchan-flat,kouchan-gauge-inv,Bruni-Gualtieri-Sopuerta}.
In particular, the procedure to find gauge invariant variables for
higher-order perturbations on a generic background spacetime was
proposed by the present author\cite{kouchan-gauge-inv} assuming that
we have already known the procedure to find gauge invariant
variables for a linear-order metric perturbations.
The contents of this paper are based on this proposal.
The main purpose of this paper is to present some formulae of
the second-order perturbative curvatures within the
two-parameter perturbation theory that are useful in some
physical applications.
When we derive these formulae, we follow the general framework
of the second-order gauge invariant perturbation theory on an
arbitrary background spacetime.
This framework is originally proposed by Stewart et
al.\cite{J.M.Stewart-M.Walker11974} and developed by Bruni et
al.\cite{Bruni-Gualtieri-Sopuerta,M.Bruni-S.Soonego-CQG1997}, and
the present author\cite{kouchan-gauge-inv}. 
These perturbative curvatures have the same form as those given
in the definitions of gauge invariant variables for arbitrary
perturbative fields which are proposed in a previous
paper\cite{kouchan-gauge-inv}. 
As in that paper, we do not make any specific assumption
regarding the background spacetime and the physical meaning of
the two-parameter family. 
Because we make no assumption concerning the background
spacetime, this framework has a wide area of applications.


The organization of this paper is as follows.
In \S\ref{sec:gauge-freedom-of-perturbations}, we review the general
framework of the second order gauge invariant perturbation theory.
We mainly review the one-parameter perturbation theory.
We emphasize that the review in
\S\ref{sec:gauge-freedom-of-perturbations} of this paper is based on 
the idea of Stewart et al.\cite{J.M.Stewart-M.Walker11974} and the
development carried out by Bruni et
al.\cite{Bruni-Gualtieri-Sopuerta,M.Bruni-S.Soonego-CQG1997}.
In \S\ref{sec:Formulae-of-perturbative-curvatures}, we present some
formulae for the second-order perturbative curvatures within the
two-parameter perturbation theory.
We also give a derivation of these formulae.
The final section, \S\ref{sec:summary}, is devoted to a summary and
brief discussion of physical situations to which this framework of
higher-order perturbation theory should be applied.
We employ the notation of our previous paper\cite{kouchan-gauge-inv}
and use the abstract index notation\cite{Wald-book}.


\section{Gauge degree of freedom in perturbation theory}
\label{sec:gauge-freedom-of-perturbations}


In this section, we briefly review the gauge degree of freedom
in general relativistic perturbations.
This was originally discussed by Stewart et
al.\cite{J.M.Stewart-M.Walker11974}. 
To explain the {\it gauge degree of freedom} in perturbation
theories, we have to keep in mind what we are doing when we
consider perturbations.
We first comment on the intuitive explanation of the gauge
degree of freedom in \S\ref{sec:what-is-gauge}.
Next, in \S\ref{sec:mathematical-formulation}, we review
the more precise mathematical formulation of the perturbations
in the theories with general covariance. 
The explanation given here is based on the works of Bruni et
al.\cite{M.Bruni-S.Soonego-CQG1997}, which represent
extensions of the idea of Stewart et al..
When we consider the perturbations in the theory with general
covariance, we have to exclude these gauge degrees of freedom in
perturbations.
To accomplish this, gauge invariant quantities of the
perturbations are useful, and these are regarded as physically
meaningful quantities. 
In \S\ref{sec:gauge-invariant-variables}, based on the
mathematical preparation given in
\S\ref{sec:mathematical-formulation}, we review the procedure to
find gauge invariant quantities of perturbations, which was
developed by the present author\cite{kouchan-gauge-inv}.


\subsection{What is ``gauge''?}
\label{sec:what-is-gauge}


\begin{wrapfigure}{r}{66mm}
    \includegraphics[width=66mm]{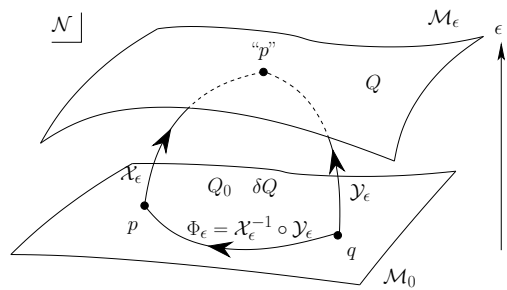}
    \caption{``gauge choice'' in one-parameter perturbation.}
    \label{fig:figure1}
\end{wrapfigure}


In perturbation theory, we always treat two spacetime
manifolds. 
One is the physical spacetime, which we attempt to describe by
perturbations, and the other is the background spacetime, which
we prepare for perturbative analyses. 
Let us denote the physical spacetime by $({\cal M},\bar{g}_{ab})$ and
the background spacetime by $({\cal M}_{0},g_{ab})$.
Keeping in mind these two spacetime manifolds, let us formally
denote the spacetime metric and the other physical tensor fields on
the physical spacetime ${\cal M}$ by $Q$. 
As the perturbation of the physical variable $Q$, we always write
equations of the form 
\begin{equation}
  \label{eq:variable-symbolic-perturbation}
  Q(``p\mbox{''}) = Q_{0}(p) + \delta Q(p).
\end{equation}
Usually, this equation is regarded as the relation between the
physical variable $Q$ and its background value $Q_{0}$ of the same
field, or simply as the definition of the deviation $\delta Q$
of $Q$ from its background value, $Q_{0}$. 
In fact, through the equation
(\ref{eq:variable-symbolic-perturbation}), we have implicitly
assigned a correspondence between points of the physical and the
background spacetime, since this equation gives a relation
between field variables $Q$, $Q_{0}$ and $\delta Q$.
More specifically, $Q(``p\mbox{''})$ in the left-hand side of
Eq.~(\ref{eq:variable-symbolic-perturbation}) is a field on the
physical spacetime ${\cal M}$, and $``p\mbox{''}\in{\cal M}$.
On the other hand, we should regard the background value
$Q_{0}(p)$ of $Q(``p\mbox{''})$ and its deviation $\delta Q(p)$
from $Q_{0}(p)$, which are on the right-hand side of
Eq.~(\ref{eq:variable-symbolic-perturbation}), as fields on 
the background spacetime ${\cal M}_{0}$ and $p\in{\cal M}_{0}$. 
Because Eq.~(\ref{eq:variable-symbolic-perturbation}) is for field
variables, it implicitly states that the points
$``p\mbox{''}\in{\cal M}$ and $p\in{\cal M}_{0}$ are same. 
This is an implicit assumption of the existence of a map 
${\cal M}_{0}\rightarrow{\cal M}$ $:$  
$p\in{\cal M}_{0}\mapsto ``p\mbox{''}\in{\cal M}$, which is usually
called a ``{\it gauge choice}'' in perturbation
theory\cite{J.M.Stewart-M.Walker11974}. 
Clearly, this is more than the usual assignment of coordinate
labels to points within the single spacetime.


It is important to note that the correspondence between points
on each ${\cal M}_{\epsilon}$, which established by
such a relation as Eq.~(\ref{eq:variable-symbolic-perturbation}),
is not unique to the theory in which general covariance is imposed. 
Rather, Eq.~(\ref{eq:variable-symbolic-perturbation})
involves the degree of freedom corresponding to the choice of
the map ${\cal X}$ $:$ ${\cal M}_{0}\rightarrow{\cal M}$ (the choice
of the point identification map ${\cal M}_{0}\rightarrow{\cal M}$). 
This is called the {\it gauge degree of freedom}.
Further, such a degree of freedom always exists in the
perturbations of a theory in which we impose general covariance,
unless, there is a preferred coordinate system in the theory
and we naturally introduce this coordinate system onto both 
${\cal M}_{0}$ and ${\cal M}$.
Then, we can choose the identification map ${\cal X}$ using this
coordinate system.
However, there is no such coordinate system, due to general
covariance, and we have no guiding principle to choose the
identification map ${\cal X}$.
Therefore, we may identify $``p\mbox{''}\in{\cal M}$ with 
$q\in{\cal M}_{0}$ ($q\neq p$) instead of $p\in{\cal M}_{0}$. 
(See Fig.~\ref{fig:figure1}.)
A gauge transformation is simply a change of the map ${\cal X}$.


\subsection{More precise formulation of perturbations}
\label{sec:mathematical-formulation}


In this section, we review the more precise formulation concerning
about ``{\it gauge degree of freedom}'' based on the above
understanding of ``{\it gauges}''
\cite{J.M.Stewart-M.Walker11974,M.Bruni-S.Soonego-CQG1997}.
We mainly review the one-parameter perturbation theory in
\S\ref{sec:one-parameter-case} and comment on the results in
two-parameter perturbation theory in \S\ref{sec:two-parameter-case}.
The essential part of the multi-parameter perturbations is
completely similar to the one-parameter
case\cite{kouchan-gauge-inv,M.Bruni-S.Soonego-CQG1997}.
Details can be seen in the papers by Bruni et
al.\cite{M.Bruni-S.Soonego-CQG1997} and by the present
author\cite{kouchan-gauge-inv}.


\subsubsection{One-parameter perturbation theory}
\label{sec:one-parameter-case}


We denote the perturbation parameter by $\epsilon$, and we
consider the $m+1$-dimensional manifold 
${\cal N}={\cal M}\times\MR$, where $m=\dim{\cal M}$ and
$\epsilon\in\MR$, as depicted in Fig.~\ref{fig:figure1}. 
By this construction, the manifold ${\cal N}$ is foliated by
$m$-dimensional submanifolds ${\cal M}_{\epsilon}$ that are
diffeomorphic to the physical spacetime ${\cal M}$.
The background ${\cal M}_{0}=\left.{\cal N}\right|_{\epsilon=0}$
and the physical spacetime
${\cal M}={\cal M}_{\epsilon}=\left.{\cal N}\right|_{\MR=\epsilon}$
are also submanifolds embedded in ${\cal N}$.
Each point on ${\cal N}$ is identified by a pair $(p,\epsilon)$,
where $p\in{\cal M}_{\epsilon}$, and each point on the background
spacetime ${\cal M}_{0}$ in ${\cal N}$ is identified by $\epsilon=0$.
The manifold ${\cal N}$ has a natural differentiable structure
consisting of the direct product of ${\cal M}$ and $\MR$.
By this construction, the perturbed spacetimes ${\cal M}_{\epsilon}$ 
for each $\epsilon$ must have the same differential structure.
In other words, we require that perturbations be continuous in the
sense that $({\cal M},\bar{g}_{ab})$ and $({\cal M}_{0},g_{ab})$ are
connected by a continuous curve within the extended spacetime
${\cal N}$. 
Hence, the changes of the differential structure resulting from the
perturbation, for example the formation of singularities and singular
perturbations in the sense of fluid mechanics, are excluded from
the study carried out in this paper.


Let us consider the set of field equations 
\begin{equation}
  \label{eq:field-eq-for-Q}
  {\cal E}[Q_{\epsilon}] = 0
\end{equation}
on the physical spacetime ${\cal M}_{\epsilon}$ for the physical
variables $Q_{\epsilon}$ on ${\cal M}_{\epsilon}$. 
The field equation (\ref{eq:field-eq-for-Q}) formally represents 
the Einstein equation for the metric on ${\cal M}_{\epsilon}$ 
and the equations for matter fields on ${\cal M}_{\epsilon}$. 
If a tensor field $Q_{\epsilon}$ is given on each 
${\cal M}_{\epsilon}$, $Q_{\epsilon}$ is automatically extended to a 
tensor field on ${\cal N}$ by $Q(p,\epsilon) := Q_{\epsilon}(p)$, 
where $p\in{\cal M}_{\epsilon}$. 
In this extension, the field equation (\ref{eq:field-eq-for-Q}) 
is regarded as the equation on the extended manifold ${\cal N}$. 
Thus, we have extended an arbitrary tensor field and field equations 
(\ref{eq:field-eq-for-Q}) on each ${\cal M}_{\epsilon}$ to those on 
the extended manifold ${\cal N}$.


Tensor fields on ${\cal N}$ obtained by the above construction
are necessarily ``tangent'' to each ${\cal M}_{\epsilon}$, i.e.,
their normal component to each ${\cal M}_{\epsilon}$ identically
vanishes.
To consider the basis of the tangent space of ${\cal N}$, we introduce
the normal form of each ${\cal M}_{\epsilon}$ in ${\cal N}$ and
its dual.
These are denoted by $(d\epsilon)_{a}$ and 
$(\partial/\partial\epsilon)^{a}$, respectively, and they
satisfy 
\begin{equation}
  (d\epsilon)_{a} \left(\frac{\partial}{\partial\epsilon}\right)^{a} = 1.
  \label{eq:depsilon-normalization}
\end{equation}
The form $(d\epsilon)_{a}$ and its dual
$(\partial/\partial\epsilon)^{a}$ are normal to any tensor field
that is extended from the tangent space on each 
${\cal M}_{\epsilon}$ by the above construction.
The set consisting of $(d\epsilon)_{a}$, $(\partial/\partial\epsilon)^{a}$ 
and the basis of the tangent space on each ${\cal M}_{\epsilon}$ is
regarded as the basis of the tangent space of ${\cal N}$.


To define the perturbation of an arbitrary tensor field $Q$, we
compare $Q$ on the physical spacetime ${\cal M}_{\epsilon}$ with
$Q_{0}$ on the background spacetime, and it is necessary to
identify the points of ${\cal M}_{\epsilon}$ with those of
${\cal M}_{0}$.
This point identification map is the so-called {\it gauge choice} 
in the context of perturbation theories, as mentioned in 
\S\ref{sec:what-is-gauge}.
The gauge choice is accomplished by assigning a diffeomorphism
${\cal X}_{\epsilon}$ $:$ ${\cal N}$ $\rightarrow$ ${\cal N}$
such that ${\cal X}_{\epsilon}$ $:$ ${\cal M}_{0}$ $\rightarrow$
${\cal M}_{\epsilon}$.
Following the paper of Bruni et al.\cite{M.Bruni-S.Soonego-CQG1997},
we introduce a gauge choice ${\cal X}_{\epsilon}$ as one of the
one-parameter groups of diffeomorphisms that satisfy the property 
\begin{equation}
  {\cal X}_{\epsilon_{1}+\epsilon_{2}}
  = {\cal X}_{\epsilon_{1}}\circ{\cal X}_{\epsilon_{2}}
  = {\cal X}_{\epsilon_{2}}\circ{\cal X}_{\epsilon_{1}}.
  \label{eq:one-parameter-group-def-prop}
\end{equation}
This one-parameter group of diffeomorphisms is generated by the
vector field ${}^{{\cal X}}\!\eta^{a}_{(\epsilon)}$.
This vector field ${}^{{\cal X}}\!\eta^{a}_{(\epsilon)}$, which
we call a {\it generator}, is defined by the action of the
corresponding pull-back ${\cal X}_{\epsilon}^{*}$ for a generic tensor
field $Q$ on ${\cal M}\times\MR$ :
\begin{equation}
  {\pounds}_{{}^{{\cal X}}\!\eta_{(\epsilon)}} Q := 
  \lim_{\epsilon\rightarrow 0} \frac{{\cal X}_{\epsilon}^{*}Q - Q}{\epsilon},
\end{equation}
and it is decomposed as
\begin{equation}
  {}^{\cal X}\!\eta^{a}_{(\epsilon)} =:
  \left(\frac{\partial}{\partial\epsilon}\right)^{a}
  + \theta^{a},
  \quad
  \theta^{a}(d\epsilon)_{a} = 0,
  \quad
  {\pounds}_{\frac{\partial}{\partial\epsilon}}\theta^{a} = 0.
  \label{eq:vector-theta-def}
\end{equation}
The third condition in (\ref{eq:vector-theta-def}) is imposed
merely for simplicity.
Except for the conditions in (\ref{eq:vector-theta-def}), we may
regard  $\theta^{a}$ as an arbitrary vector field on
${\cal M}_{\epsilon}$ (not on ${\cal N}$); i.e., the
arbitrariness of the gauge choice is given by that of the vector
field $\theta^{a}$.


The Taylor expansion of the pull-back ${\cal X}_{\epsilon}^{*}Q$
is given by 
\begin{equation}
  {\cal X}_{\epsilon}^{*}Q
  = 
  \sum_{k=0}^{\infty} \frac{\epsilon^{k}}{k!} 
  \left[
    \frac{\partial^{k}}{\partial\epsilon^{k}} {\cal X}_{\epsilon}^{*}Q
  \right]_{\epsilon=0}
  = 
  \sum_{k=0}^{\infty} \frac{\epsilon^{k}}{k!} 
  {\pounds}_{{}^{{\cal X}}\!\eta_{(\epsilon)}}^{k}Q. 
  \label{eq:Taylor-expansion-of-calX}
\end{equation}
Once the definition of the pull-back of the gauge choice 
${\cal X}_{\epsilon}$ is given, the perturbation 
$\Delta^{\cal X}Q_{\epsilon}$ of a tensor field $Q$ under the gauge
choice ${\cal X}_{\epsilon}$ is simply defined as
\begin{equation}
  \label{eq:Bruni-34}
  \Delta^{\cal X}Q_{\epsilon} :=
  \left.{\cal X}^{*}_{\epsilon}Q\right|_{{\cal M}_{0}} - Q_{0}.
\end{equation}
We note that all the variables in this definition are defined on
${\cal M}_{0}$.
The first term on the right-hand side of (\ref{eq:Bruni-34}) can
be Taylor-expanded as 
\begin{equation}
  \label{eq:Bruni-35}
  \left.{\cal X}^{*}_{\epsilon}Q_{\epsilon}\right|_{{\cal M}_{0}}
  =
  \sum^{\infty}_{k=0} \frac{\epsilon^{k}}{k!}
  {}^{(k)}_{\;\cal X}\!Q.
\end{equation}
Equations (\ref{eq:Bruni-34}) and (\ref{eq:Bruni-35}) define the
perturbation of $O(k)$ of a physical variable $Q_{\epsilon}$
under the gauge choice ${\cal X}$ and its background value
${}^{(0)}_{\;\cal X}\!Q=Q_{0}$.
Through Eqs. (\ref{eq:Taylor-expansion-of-calX}) and
(\ref{eq:Bruni-35}), each order perturbation 
${}^{(k)}_{\;\cal X}\!Q$ under the gauge choice 
${\cal X}_{\epsilon}$ is given by 
\begin{eqnarray}
  {}^{(k)}_{\;\cal X}\!Q = 
  \left.{\pounds}_{{}^{\cal X}\!\eta}^{k} Q\right|_{{\cal M}_{0}}.
\end{eqnarray}


The above understanding of the gauge choice and perturbations
naturally leads to the {\it gauge transformation rules} between
{\it different gauge choices} and the concept of 
{\it gauge invariance} as follows.


Suppose that ${\cal X}_{\epsilon}$ and ${\cal Y}_{\epsilon}$ are
two one-parameter groups of diffeomorphisms with the generators 
${}^{\cal X}\eta^{a}$ and ${}^{\cal Y}\eta^{a}$ on ${\cal N}$,
respectively, i.e., ${\cal X}_{\epsilon}$ and 
${\cal Y}_{\epsilon}$ are two gauge choices.
These generators are decomposed in the same manner as
Eqs.~(\ref{eq:vector-theta-def}):
\begin{eqnarray}
  {}^{\cal X}\!\eta^{a}
  = \left(\frac{\partial}{\partial\epsilon}\right)^{a} + \theta^{a},
  \quad
  {}^{\cal Y}\!\eta^{a}
  = \left(\frac{\partial}{\partial\epsilon}\right)^{a} + \iota^{a}.
\end{eqnarray}
The integral curves of each ${}^{\cal X}\!\eta^{a}$ and
${}^{\cal Y}\!\eta^{a}$ in ${\cal N}$ are the orbit of the action of
the gauge choices ${\cal X}_{\epsilon}$ and ${\cal Y}_{\epsilon}$,
respectively.
Since the generators ${}^{\cal X}\!\eta^{a}$ and
${}^{\cal Y}\!\eta^{a}$ are transverse to each ${\cal M}_{\epsilon}$
everywhere on ${\cal N}$, the integral curves of these vector field
intersect with each ${\cal M}_{\epsilon}$.
Therefore, points lying on the same integral curve of either of the two are
to be regarded as {\it the same point} within the respective gauges
(see Fig.~\ref{fig:figure1}).
Hence, ${\cal X}_{\epsilon}$ and ${\cal Y}_{\epsilon}$ are both point
identification maps.
When $\theta^{a}\neq\iota^{a}$, these point identification maps are
regarded as {\it two different gauge choices}.


Suppose that ${\cal X}_{\epsilon}$ and ${\cal Y}_{\epsilon}$ are
two different gauge choices which are generated by the vector
fields ${}^{\cal X}\!\eta^{a}$ and ${}^{\cal Y}\!\eta^{a}$,
respectively. 
These gauge choices also pull back a generic tensor field $Q$ on
${\cal N}$ to two other tensor fields, ${\cal X}_{\epsilon}^{*}Q$ and 
${\cal Y}_{\epsilon}^{*}Q$, for any given value of $\epsilon$.
In particular, on ${\cal M}_{0}$, we now have three
tensor fields associated with a tensor field $Q$; i.e., one is the
background value $Q_{0}$ of $Q$, and the other two are the
pulled back variables of $Q$ from ${\cal M}_{\epsilon}$ to
${\cal M}_{0}$ by the two different gauge choices,
\begin{eqnarray}
  {}^{\cal X}Q_{\epsilon} &:=&
  \left.{\cal X}^{*}_{\epsilon}Q\right|_{{\cal M}_{0}}
  = \sum^{\infty}_{k=0}
  \frac{\epsilon^{k}}{k!}
  {}^{(k)}_{\;{\cal X}}\!Q
  = Q_{0} + \Delta^{\cal X}Q_{\epsilon},
  \label{eq:Bruni-39-one}
  \\
  {}^{\cal Y}Q_{\epsilon} &:=&
  \left.{\cal Y}^{*}_{\epsilon}Q\right|_{{\cal M}_{0}}
  = \sum^{\infty}_{k=0}
  \frac{\epsilon^{k}}{k!}
  {}^{(k)}_{\;\cal Y}\!Q
  = Q_{0} + \Delta^{\cal Y}Q_{\epsilon}.
  \label{eq:Bruni-40-one}
\end{eqnarray}
Here, we have used Eqs.~(\ref{eq:Bruni-34}) and (\ref{eq:Bruni-35}).
Because ${\cal X}_{\epsilon}$ and ${\cal Y}_{\epsilon}$ are gauge
choices that map the background spacetime ${\cal M}_{0}$ into the
physical spacetime ${\cal M}_{\epsilon}$, 
${}^{\cal X}Q_{\epsilon}$ and 
${}^{\cal Y}Q_{\epsilon}$ are the representations on 
${\cal M}_{0}$ of the perturbed tensor field $Q$ in the two
different gauges. 
The quantities ${}^{(k)}_{\;\cal X}\!Q$ and 
${}^{(k)}_{\;\cal Y}\!Q$ in Eqs.~(\ref{eq:Bruni-39-one}) and
(\ref{eq:Bruni-40-one}) are the perturbations of $O(k)$ in the
gauges ${\cal X}$ and ${\cal Y}$, respectively.


Now, we consider the concept of {\it gauge invariance}.
Following the paper of Bruni et al.\cite{Bruni-Gualtieri-Sopuerta}, 
we consider the concept of {\it gauge invariance up to order $n$}.
We say that $Q$ is {\it gauge invariant up to order $n$}
iff for any two gauges ${\cal X}$ and ${\cal Y}$
\begin{equation}
  {}^{(k)}_{\;\cal X}\!Q = {}^{(k)}_{\;\cal Y}\!Q \quad 
  \forall k, \quad \mbox{with} \quad k<n.
\end{equation}
From this definition, we can prove that the $n$th-order
perturbation of a tensor field $Q$ is gauge invariant up to order
$n$ iff in a given gauge ${\cal X}$ we have ${\pounds}_{\xi}
{}^{(k)}_{\;\cal X}\!Q = 0$ for any vector field $\xi^{a}$ defined
on ${\cal M}_{0}$ and for any $k<n$.
As a consequence, the $n$th-order perturbation of a tensor field $Q$
is gauge invariant up to order $n$ iff $Q_{0}$ and all its
perturbations of lower than $n$th order are, in any gauge, either
vanishing or constant scalars, or a combination of Kronecker deltas
with constant coefficients\cite{Bruni-Gualtieri-Sopuerta,J.M.Stewart-M.Walker11974,M.Bruni-S.Soonego-CQG1997}.


In general, the representation ${}^{\cal X}Q_{\epsilon}$ on
${\cal M}_{0}$ of the perturbed variable $Q$ on ${\cal M}_{\epsilon}$
depends on the gauge choice ${\cal X}_{\epsilon}$.
If we apply a different gauge choice, the representation of
$Q_{\epsilon}$ on ${\cal M}_{0}$ may change. 
Recalling that the gauge choice ${\cal X}$ is a point
identification map from ${\cal M}_{0}$ to  ${\cal M}_{\epsilon}$
(see Fig.~\ref{fig:figure1}), the change of the gauge choice
from ${\cal X}_{\epsilon}$ to ${\cal Y}_{\epsilon}$ is
represented by the diffeomorphism
\begin{equation}
  \label{eq:diffeo-def-from-Xinv-Y}
  \Phi_{\epsilon} :=
  ({\cal X}_{\epsilon})^{-1}\circ{\cal Y}_{\epsilon}.
\end{equation}
This diffeomorphism $\Phi_{\epsilon}$ is the map
$\Phi_{\epsilon}$ $:$ ${\cal M}_{0}$ $\rightarrow$
${\cal M}_{0}$ for each value of $\epsilon\in\MR$.
As shown in Fig.~\ref{fig:figure1}, the diffeomorphism
$\Phi_{\epsilon}$ changes the point identification, as expected
from the understanding of the gauge choice discussed in 
\S\ref{sec:what-is-gauge}.
Therefore, the diffeomorphism $\Phi_{\epsilon}$ is regarded as
the gauge transformation $\Phi_{\epsilon}$ $:$ ${\cal X}_{\epsilon}$
$\rightarrow$ ${\cal Y}_{\epsilon}$.


The gauge transformation $\Phi_{\epsilon}$ induces a pull-back
from the representation ${}^{\cal X}Q_{\epsilon}$ of the
perturbed tensor field in the gauge choice ${\cal X}_{\epsilon}$
to the representation ${}^{\cal Y}Q_{\epsilon}$ in the gauge
choice ${\cal Y}_{\epsilon}$.
Actually, the tensor fields ${}^{\cal X}Q_{\epsilon}$ and
${}^{\cal Y}Q_{\epsilon}$, which are defined on ${\cal M}_{0}$,
are connected by the linear map $\Phi^{*}_{\epsilon}$ as
\begin{eqnarray}
  {}^{\cal Y}Q_{\epsilon}
  &=&
  \left.{\cal Y}^{*}_{\epsilon}Q\right|_{{\cal M}_{0}}
  =
  \left.\left(
      {\cal Y}^{*}_{\epsilon}
      \left({\cal X}_{\epsilon}
      {\cal X}_{\epsilon}^{-1}\right)^{*}Q\right)
  \right|_{{\cal M}_{0}}
  \nonumber\\
  &=&
  \left.
    \left(
      {\cal X}^{-1}_{\epsilon}
      {\cal Y}_{\epsilon}
    \right)^{*}
    \left(
      {\cal X}^{*}_{\epsilon}Q
    \right)
  \right|_{{\cal M}_{0}}
  =  \Phi^{*}_{\epsilon} {}^{\cal X}Q_{\epsilon}.
  \label{eq:Bruni-45-one}
\end{eqnarray}
According to generic arguments concerning the Taylor expansion
of the pull-back of a tensor field on the same
manifold\cite{kouchan-gauge-inv,M.Bruni-S.Soonego-CQG1997},
the gauge transformation $\Phi^{*}_{\epsilon} {}^{\cal X}Q_{\epsilon}$
should be given by the form
\begin{eqnarray}
  \Phi^{*}_{\epsilon}{}^{\cal X}\!Q = {}^{\cal X}\!Q
  + \epsilon {\pounds}_{\xi_{1}} {}^{\cal X}\!Q
  + \frac{\epsilon^{2}}{2} \left\{
    {\pounds}_{\xi_{2}} + {\pounds}_{\xi_{1}}^{2}
  \right\} {}^{\cal X}\!Q
  + O(\epsilon^{3}),
  \label{eq:Bruni-46-one} 
\end{eqnarray}
where the vector fields $\xi_{1}^{a}$ and $\xi_{2}^{a}$ are the
generators of the gauge transformation $\Phi_{\epsilon}$.


Comparing the representation (\ref{eq:Bruni-46-one}) of the expansion
in terms of the generators $\xi_{p}^{a}$ of the pull-back
$\Phi_{\epsilon}^{*}{}^{\cal X}\!Q$ and that in terms of the generators
${}^{\cal X}\eta_{(\epsilon)}^{a}$ and
${}^{\cal Y}\eta_{(\epsilon)}^{a}$ of the pull-back
${\cal Y}^{*}_{\epsilon}\circ\left({\cal X}_{\epsilon}^{-1}\right)^{*}Q$
($=\Phi_{\epsilon}^{*}{}^{\cal X}\!Q$), we easily find explicit
expressions for the generators $\xi_{p}^{a}$ of the gauge
transformation $\Phi={\cal X}^{-1}\circ{\cal Y}$ in terms of the
generators ${}^{\cal X}\eta_{(\epsilon)}^{a}$ and
${}^{\cal Y}\eta_{(\epsilon)}^{a}$ of the gauge choices.
Further, because the gauge transformation $\Phi_{\epsilon}$ is a
map within the background spacetime ${\cal M}_{0}$, the
generator should be given as vector fields on ${\cal M}_{0}$.
The explicit expression of the generators $\xi_{p}^{a}$ in terms of
the components of the generators of the gauge choices is given
in some papers\cite{kouchan-gauge-inv,M.Bruni-S.Soonego-CQG1997}.


We can now derive the relation between the perturbations in the two
different gauges.
Up to second order, these relations are derived by substituting
(\ref{eq:Bruni-39-one}) and (\ref{eq:Bruni-40-one}) into
(\ref{eq:Bruni-46-one}): 
\begin{eqnarray}
  \label{eq:Bruni-47-one} 
  {}^{(1)}_{\;{\cal Y}}\!Q - {}^{(1)}_{\;{\cal X}}\!Q &=& 
  {\pounds}_{\xi_{1}}Q_{0}, \\
  \label{eq:Bruni-49-one} 
  {}^{(2)}_{\;\cal Y}\!Q - {}^{(2)}_{\;\cal X}\!Q &=& 
  2 {\pounds}_{\xi_{(1)}} {}^{(1)}_{\;\cal X}\!Q 
  +\left\{{\pounds}_{\xi_{(2)}}+{\pounds}_{\xi_{(1)}}^{2}\right\} Q_{0}.
\end{eqnarray}
These results are, of course, consistent with the concept of
gauge invariance up to order $n$, as introduced above.
Inspecting these gauge transformation rules, we can define the
gauge invariant variables.


\subsubsection{Two-parameter perturbation theory}
\label{sec:two-parameter-case}


Here, we briefly review the two-parameter case.
We denote the two parameters for the perturbation by $\epsilon$
and $\lambda$.
In this case, we have to consider the extended manifold 
${\cal N}={\cal M}\times\MR^{2}$, instead of 
${\cal N}={\cal M}\times\MR$ used in the one-parameter case,
where $(\epsilon,\lambda)\in\MR^{2}$.
As in the one-parameter case, the gauge choice 
${\cal X}_{\epsilon,\lambda}$ is a point identification map
${\cal X}_{\epsilon,\lambda}:{\cal M}_{0}\rightarrow{\cal M}$ on
${\cal N}$.
This gauge choice ${\cal X}_{\epsilon,\lambda}$ has the property 
\begin{equation}
  {\cal X}_{\epsilon_{1},\lambda_{1}}\circ{\cal X}_{\epsilon_{2},\lambda_{2}}
  = 
  {\cal X}_{\epsilon_{1}+\epsilon_{2},\lambda_{1}+\lambda_{2}} 
  \quad
  \lambda_{1}, \lambda_{2}, \epsilon_{1},\epsilon_{2} \in \MR.
  \label{eq:two-parameter-gauge-choice-property}
\end{equation}
This property implies that 
\begin{equation}
  {\cal X}_{\epsilon,\lambda} = {\cal X}_{\epsilon,0}\circ{\cal X}_{0,\lambda}
  = {\cal X}_{0,\lambda}\circ{\cal X}_{\epsilon,0},
  \label{eq:two-parameter-gauge-choice-fixed}
\end{equation}
where ${\cal X}_{\epsilon,0}$ and ${\cal X}_{0,\lambda}$ are two
one-parameter groups of diffeomorphisms defined by the property
Eq.~(\ref{eq:one-parameter-group-def-prop}).


We denote the generators of ${\cal X}_{\epsilon,0}$ and 
${\cal X}_{0,\lambda}$ by ${}^{\cal X}\eta^{a}_{(\epsilon)}$ and
${}^{\cal X}\eta^{a}_{(\lambda)}$, respectively.
We also introduce the basis $(\partial/\partial\epsilon)^{a}$,
$(d\epsilon)_{a}$, $(\partial/\partial\lambda)^{a}$, $(d\lambda)_{a}$
as vector fields on ${\cal N}$, which satisfy conditions similar
to those in (\ref{eq:depsilon-normalization})\cite{kouchan-gauge-inv}.
Using these basis, the generators ${}^{\cal X}\eta^{a}_{(\epsilon)}$
(${}^{\cal X}\eta^{a}_{(\lambda)}$) of the one-parameter group
of diffeomorphisms ${\cal X}_{\epsilon,0}$ (${\cal X}_{0,\lambda}$)
is decomposed in the same manner as in Eq.~(\ref{eq:vector-theta-def}).
The property (\ref{eq:two-parameter-gauge-choice-fixed}) is
expressed by  
\begin{eqnarray}
  [{}^{\cal X}\eta_{(\epsilon)},{}^{\cal X}\eta_{(\lambda)}]^{a} = 0
\end{eqnarray}
in terms of these generators. 
The Taylor expansion of the pull-back 
${\cal X}_{\epsilon,\lambda}^{*}Q$ is given by 
\begin{eqnarray}
  {\cal X}_{\epsilon,\lambda}^{*} Q &=& \sum_{k,k'=0}^{\infty}
  \frac{\lambda^{k}\epsilon^{k'}}{k!k'!} 
  \left[
    \frac{\partial^{k+k'}}{\partial\lambda^{k}\partial\epsilon^{k'}} 
    {\cal X}_{\epsilon,\lambda}^{*} Q
  \right]_{\lambda=\epsilon=0}
  \\
  &=& \sum_{k,k'=0}^{\infty} \frac{\lambda^{k}\epsilon^{k'}}{k!k'!} 
  \pounds_{{}^{\cal X}\eta_{(\lambda)}}^{k}
  \pounds_{{}^{\cal X}\eta_{(\epsilon)}}^{k'}
  Q.
  \label{eq:two-parameter-gauge-Taylor}
\end{eqnarray}
The perturbation $\Delta_{0}^{\cal X}Q_{\epsilon,\lambda}$ of an
arbitrary tensor field $Q$ in terms of the gauge choice 
${\cal X}_{\epsilon,\lambda}$ is given by 
\begin{eqnarray}
  \Delta_{0}^{\cal X} Q_{\epsilon,\lambda} :=
  \left.{\cal X}_{\epsilon,\lambda}^{*}Q\right|_{{\cal M}_{0}} - Q_{0},
  \quad
  \left.{\cal X}_{\epsilon,\lambda}^{*} Q\right|_{{\cal M}_{0}} 
  = \sum_{k,k'=0}^{\infty}
  \frac{\lambda^{k}\epsilon^{k'}}{k!k'!} \;\;
  {}_{\quad\cal X}^{(k,k')}\!Q ,
  \label{eq:two-parameter-perturbation-def}
\end{eqnarray}
where ${}_{\quad\cal X}^{(k,k')}\!Q$ is the perturbation of order
$(k,k')$ and ${}_{\quad\cal X}^{(0,0)}\!Q=Q_{0}$.
Together with the expansion given in 
(\ref{eq:two-parameter-gauge-Taylor}) and
(\ref{eq:two-parameter-perturbation-def}), each order
perturbation ${}^{(k,k')}_{\quad\cal X}Q$ with the gauge choice
${\cal X}_{\epsilon,\lambda}$ is given by 
\begin{eqnarray}
  {}^{(k,k')}_{\quad\cal X}Q = 
  \left.\pounds_{{}^{\cal X}\eta_{(\lambda)}}^{k}
  \pounds_{{}^{\cal X}\eta_{(\epsilon)}}^{k'}
  Q\right|_{{\cal M}_{0}}.
\end{eqnarray}


The concept of different gauges, a gauge transformation, gauge
invariance, and the definition of gauge transformation rules in
the two-parameter case are similar to those in the one-parameter
case. 
For this reason, we can derive the following gauge
transformation rules:\cite{kouchan-gauge-inv}
\begin{eqnarray}
  \label{eq:Bruni-47-two}
  {}^{(p,q)}_{\quad{\cal Y}}Q - {}^{(p,q)}_{\quad{\cal X}}Q &=& 
  {\pounds}_{\xi_{(p,q)}}Q_{0}
  \quad\quad\quad\quad\quad\quad\quad
  \mbox{for} \quad
  (p,q) = (1,0), (0,1),
  \\
  \label{eq:Bruni-49-two}
  {}^{(p,q)}_{\quad{\cal Y}}Q - {}^{(p,q)}_{\quad{\cal X}}Q &=& 
  2 {\pounds}_{\xi_{(\frac{p}{2},\frac{q}{2})}} 
    {}^{(\frac{p}{2},\frac{q}{2})}_{\quad{\cal X}} Q 
  +\left\{{\pounds}_{\xi_{(p,q)}} 
    + {\pounds}_{\xi_{(\frac{p}{2},\frac{q}{2})}}^{2}
  \right\} Q_{0},
  \nonumber\\
  &&
  \quad\quad\quad\quad\quad\quad\quad
  \quad\quad\quad\quad
  \mbox{for} \quad
  (p,q) = (2,0), (0,2),
  \\
  \label{eq:Bruni-50-two}
  {}^{(1,1)}_{\quad{\cal Y}}Q - {}^{(1,1)}_{\quad{\cal X}}Q &=& 
  {\pounds}_{\xi_{(1,0)}} {}^{(0,1)}_{\quad{\cal X}}Q 
  + {\pounds}_{\xi_{(0,1)}} {}^{(1,0)}_{\quad{\cal X}}Q 
  \nonumber\\
  && 
  + \left\{{\pounds}_{\xi_{(1,1)}} 
    + \frac{1}{2} {\pounds}_{\xi_{(1,0)}}{\pounds}_{\xi_{(0,1)}}
    + \frac{1}{2} {\pounds}_{\xi_{(0,1)}}{\pounds}_{\xi_{(1,0)}}
  \right\} Q_{0},
\end{eqnarray}
where the $\xi_{(p,q)}^{a}$ are the generators for gauge
transformation $\Phi_{\epsilon,\lambda}:=\left({\cal X}_{\epsilon,\lambda}\right)^{-1}\circ{\cal Y}_{\epsilon,\lambda}$.


In this paper, we treat these gauge transformation rules of
two-parameter perturbation theory as mentioned in the
introduction (\S\ref{sec:intro}), because the one-parameter case
considered above can be treated as a special case of this
two-parameter case.


\subsection{Gauge invariant variables}
\label{sec:gauge-invariant-variables}


Inspecting the gauge transformation rules
(\ref{eq:Bruni-47-two})--(\ref{eq:Bruni-50-two}), we can define
the gauge invariant variables for a metric perturbation and for
arbitrary matter fields.
Employing the idea of gauge invariance up to order $n$ for
$n$th-order perturbations\cite{M.Bruni-S.Soonego-CQG1997}, we
proposed the procedure to construct gauge invariant variables of
higher-order perturbations\cite{kouchan-gauge-inv}.
This proposal is as follows.
First, we construct gauge invariant variables for the metric
perturbation.
Then, we define the gauge invariant variables for perturbations
of an arbitrary field, excluding perturbations of the metric.
The procedure to find the gauge invariant part of a higher-order 
perturbation is a simple extension of that for linear-order
perturbations.


To consider the metric perturbation, we expand the metric on the
physical spacetime ${\cal M}$, which is pulled back to the
background spacetime ${\cal M}_{0}$ using a gauge choice in the
form given in (\ref{eq:two-parameter-perturbation-def}): 
\begin{eqnarray}
  {\cal X}^{*}_{\epsilon,\lambda}\bar{g}_{ab} &=& 
  \sum_{k',k=0}^{\infty}
  \frac{\epsilon^{k}\lambda^{k'}}{k!k'!} 
  {}^{(k,k')}_{\;\;\;\;\;{\cal X}}h_{ab} 
  \\
  &=& g_{ab} + \epsilon {}^{(1,0)}_{\;\;\;\;\;{\cal X}}h_{ab} 
  + \lambda {}^{(0,1)}_{\;\;\;\;\;{\cal X}}h_{ab}
  \nonumber\\
  && \quad\quad
  + \frac{\epsilon^{2}}{2} {}^{(2,0)}_{\;\;\;\;\;{\cal X}}h_{ab}
  + \epsilon\lambda {}^{(1,1)}_{\;\;\;\;\;{\cal X}}h_{ab}
  + \frac{\lambda^{2}}{2} {}^{(0,2)}_{\;\;\;\;\;{\cal X}}h_{ab}
  + O^{3}(\epsilon,\lambda),
  \label{eq:metric-expansion}
\end{eqnarray}
where ${}^{(0,0)}h_{ab}=g_{ab}$ is the metric on the background
spacetime ${\cal M}_{0}$.
Of course, the expansion (\ref{eq:metric-expansion}) of the metric
depends entierly on the gauge choice ${\cal X}$.
Nevertheless, we do not explicitly express the index of the
gauge choice ${\cal X}$ in an expression if there is no
possibility of confusion.


Our starting point to construct gauge invariant variables is the
assumption that {\it we already know the procedure
to find gauge invariant variables for the linear metric
perturbations.} Then, linear metric perturbations
${}^{(1,0)}h_{ab}$ (${}^{(0,1)}h_{ab}$) are decomposed as 
\begin{eqnarray}
  {}^{(p,q)}h_{ab} 
  =: {}^{(p,q)}{\cal H}_{ab} 
  + 2 \nabla_{(a}\;{}^{(p,q)}\!X_{b)}, 
  \quad
  (p,q) = (1,0), (0,1),
  \label{eq:linear-metric-decomp}
\end{eqnarray}
where ${}^{(p,q)}{\cal H}_{ab}$ and ${}^{(p,q)}X_{a}$ are the
gauge invariant and variant parts of the linear-order metric
perturbations\cite{kouchan-gauge-inv}.
Hence, under the gauge transformation (\ref{eq:Bruni-47-two}),
these are transformed as 
${}^{(p,q)}_{\quad{\cal Y}}{\cal H}_{ab}$ $-$  
${}^{(p,q)}_{\quad{\cal X}}{\cal H}_{ab}$ $=$ $0$
and
${}^{(p,q)}_{\quad{\cal Y}}X^{a}$ $-$ 
${}^{(p,q)}_{\quad{\cal X}}X^{a}$ $=$ $\xi^{a}_{(p,q)}$.


As emphasized in a previous paper\cite{kouchan-gauge-inv}, the
above assumption is quite strong and it is not trivial to carry
out the systematic decomposition (\ref{eq:linear-metric-decomp}) 
on an arbitrary background spacetime, as this procedure
depends completely on the background spacetime $({\cal M}_{0},g_{ab})$. 
However, this procedure is known in the perturbation theory on
some simple background spacetimes, for example the cosmological
perturbations of homogeneous and isotropic
universes\cite{Cosmological-Perturbations} or perturbations of 
spherically symmetric spacetimes\cite{Gerlach_Sengupta}.
Further, from a general point of view, knowledge of linear
perturbation theory is always necessary to carry out the 
second-order perturbations.
For these reasons, we start from this assumption in spite of the
fact that it is quite strong.


Once we accept this assumption, we can always find gauge
invariant variables for higher-order
perturbations\cite{kouchan-gauge-inv}. 
As shown in a previous paper\cite{kouchan-gauge-inv}, at
second-order, the metric perturbations are decomposed as 
\begin{eqnarray}
  \label{eq:H-ab-in-gauge-X-def-second-1}
  {}^{(p,q)}h_{ab}
  &=:&
  {}^{(p,q)}{\cal H}_{ab}
  + 2 {\pounds}_{{}^{(\frac{p}{2},\frac{q}{2})}X} 
  {}^{(\frac{p}{2},\frac{q}{2})}h_{ab}
  \nonumber\\
  && \quad
  + \left(
      {\pounds}_{{}^{(p,q)}X}
    - {\pounds}_{{}^{(\frac{p}{2},\frac{q}{2})}X}^{2} 
  \right)
  g_{ab},
  \quad\quad
  (p,q) = (2,0), (0,2);
  \\
  \label{eq:H-ab-in-gauge-X-def-second-2}
  {}^{(1,1)}h_{ab}
  &=:&
  {}^{(1,1)}{\cal H}_{ab}
  + {\pounds}_{{}^{(0,1)}X} {}^{(1,0)}h_{ab}
  + {\pounds}_{{}^{(1,0)}X} {}^{(0,1)}h_{ab}
  \nonumber\\
  && \quad
  + \left\{
    {\pounds}_{{}^{(1,1)}X}
    - \frac{1}{2}
      {\pounds}_{{}^{(1,0)}X}
      {\pounds}_{{}^{(0,1)}X}
    - \frac{1}{2}
      {\pounds}_{{}^{(0,1)}X}
      {\pounds}_{{}^{(1,0)}X}
  \right\} g_{ab},
\end{eqnarray}
where ${}^{(p,q)}{\cal H}_{ab}$ and ${}^{(p,q)}X_{a}$ are the
gauge invariant and variant parts of the metric perturbations
under the gauge transformation rules
(\ref{eq:Bruni-47-two})--(\ref{eq:Bruni-50-two}).


Furthermore, using the gauge variant parts ${}^{(p,q)}X_{a}$ of
metric perturbations\cite{kouchan-gauge-inv}, gauge invariant 
variables for an arbitrary field $Q$ excluding the metric are
given by\cite{kouchan-gauge-inv} 
\begin{eqnarray}
  \label{eq:matter-gauge-inv-def-1.0} 
  {}^{(p,q)}\!{\cal Q} &:=&
  {}^{(p,q)}\!Q - {\pounds}_{{}^{(p,q)}X}Q_{0},
  \quad\quad\quad\quad\quad\quad\quad
  (p,q) = (1,0), (0,1)
  , \\ 
  \label{eq:matter-gauge-inv-def-2.0} 
  {}^{(p,q)}\!{\cal Q} &:=&
  {}^{(p,q)}\!Q 
  - 2 {\pounds}_{{}^{(\frac{p}{2},\frac{q}{2})}X} 
  {}^{(\frac{p}{2},\frac{q}{2})}Q 
  \nonumber\\
  && \quad
  - \left\{
    {\pounds}_{{}^{(p,q)}X}
    -{\pounds}_{{}^{(\frac{p}{2},\frac{q}{2})}X}^{2}
  \right\} Q_{0},
  \quad\quad\quad
  (p,q) = (2,0), (0,2)
  , \\
  \label{eq:matter-gauge-inv-def-1.1} 
  {}^{(1,1)}\!{\cal Q} &:=&
  {}^{(1,1)}\!Q
  - {\pounds}_{{}^{(1,0)}X} {}^{(0,1)}\!Q
  - {\pounds}_{{}^{(0,1)}X} {}^{(1,0)}\!Q
  \nonumber\\
  && \quad
  - \left\{{\pounds}_{{}^{(1,1)}X}
    - \frac{1}{2} {\pounds}_{{}^{(1,0)}X}
                  {\pounds}_{{}^{(0,1)}X}
    - \frac{1}{2} {\pounds}_{{}^{(0,1)}X}
                  {\pounds}_{{}^{(1,0)}X}
    \right\} Q_{0}.
\end{eqnarray}
It is straightforward to confirm that the variables
${}^{(p,q)}\!{\cal Q}$ defined by
(\ref{eq:matter-gauge-inv-def-1.0})--(\ref{eq:matter-gauge-inv-def-1.1})
are gauge invariant under the gauge transformation rules
(\ref{eq:Bruni-47-two})--(\ref{eq:Bruni-50-two}). 
In this paper, we derive some formulae for second-order
perturbations of curvatures from the expansion of the metric
perturbation on a generic spacetime.
The starting point of this derivation is the decomposition
(\ref{eq:linear-metric-decomp})--(\ref{eq:H-ab-in-gauge-X-def-second-2})
of the metric perturbations in terms of the gauge invariant and variant
variables. 
As a result, we find that all formulae have forms which are
similar to those given in the definitions
(\ref{eq:matter-gauge-inv-def-1.0})--(\ref{eq:matter-gauge-inv-def-1.1})
of the gauge invariant variables for arbitrary matter fields.


\section{Formulae of perturbative curvatures}
\label{sec:Formulae-of-perturbative-curvatures}


Now, we derive the formulae for the perturbative curvatures at
each order in two parameter perturbation theory, following the
standard derivation of the perturbative curvature\cite{Wald-book}.


The starting point of the derivation is simply the definition of
the curvature $\bar{R}_{abc}^{\;\;\;\;\;\;d}$ on the physical
spacetime $({\cal M},\bar{g}_{ab})$
\begin{eqnarray}
  \left(
    \bar{\nabla}_{a}\bar{\nabla}_{b}
    - \bar{\nabla}_{b}\bar{\nabla}_{a}
  \right) \bar{\omega}_{c}
  = \bar{\omega}_{d}\bar{R}_{abc}^{\;\;\;\;\;\;d},
  \label{eq:physical-spacetime-Riemann-def}
\end{eqnarray}
where $\bar{\nabla}_{a}$ is the covariant derivative associated
with the metric $\bar{g}_{ab}$ on the physical spacetime ${\cal M}$
and $\bar{\omega}_{c}$ is an arbitrary one-form on the physical
spacetime ${\cal M}$.
We similarly define the curvature $R_{abc}^{\;\;\;\;\;\;d}$ on
the background spacetime $({\cal M}_{0},g_{ab})$,
\begin{eqnarray}
  \left(
    \nabla_{a}\nabla_{b}
    - \nabla_{b}\nabla_{a}
  \right) \omega_{c}
  = \omega_{d}R_{abc}^{\;\;\;\;\;\;d},
  \label{eq:background-spacetime-Riemann-def}
\end{eqnarray}
where $\nabla_{a}$ is the covariant derivative associated
with the metric $g_{ab}$ on the background spacetime 
${\cal M}_{0}$, and $\omega_{c}$ is an arbitrary one-form on the 
background spacetime ${\cal M}_{0}$.
Our task is to compare $\bar{R}_{abc}^{\;\;\;\;\;\;d}$
and $R_{abc}^{\;\;\;\;\;\;d}$.
To accomplish this, we have to consider the difference between
the gauge choices for the physical spacetime ${\cal M}$ and the
background spacetime ${\cal M}_{0}$ as discussed above.


To compare the Riemann curvature
(\ref{eq:physical-spacetime-Riemann-def}) of the physical spacetime
${\cal M}$ and that (\ref{eq:background-spacetime-Riemann-def})
of the background spacetime ${\cal M}_{0}$, we introduce the
derivative operator 
${\cal X}^{*}\bar{\nabla}_{a}\left({\cal X}^{-1}\right)^{*}$ 
on the background spacetime ${\cal M}_{0}$. This derivative
operator ${\cal X}^{*}\bar{\nabla}_{a}\left({\cal X}^{-1}\right)^{*}$
is the pull-back of the covariant derivative $\nabla_{a}$
associated with the metric $\bar{g}_{ab}$ on the physical
spacetime ${\cal M}$. 
The property of the derivative operator
${\cal X}^{*}\bar{\nabla}_{a}\left({\cal X}^{-1}\right)^{*}$ as the
covariant derivative is given by
\begin{equation}
  {\cal X}^{*}\bar{\nabla}_{a}
  \left(
    \left(
      {\cal X}^{-1}\right)^{*}{\cal X}^{*}\bar{g}_{ab}
  \right) = 0,
  \label{eq:property-as-covariant-derivative-on-phys-sp}
\end{equation}
where ${\cal X}^{*}\bar{g}_{ab}$ is the pull-back of the metric
on the physical spacetime ${\cal M}$, which is expanded as
Eq.~(\ref{eq:metric-expansion}). 
Through the introduction of this operator 
${\cal X}^{*}\bar{\nabla}_{a}\left({\cal X}^{-1}\right)^{*}$, we can
regard the definition of the Riemann curvature
(\ref{eq:physical-spacetime-Riemann-def}) on the physical spacetime
${\cal M}$ as an equation on the background spacetime.
Since the pull-back
${\cal X}^{*}\bar{\nabla}_{a}\left({\cal X}^{-1}\right)^{*}$ on the
background spacetime ${\cal M}_{0}$ of the covariant derivative
$\bar{\nabla}_{a}$ on the physical spacetime ${\cal M}$ is
linear, satisfies the Leibnitz rule, commutes with contraction,
is consistent with the concept of tangent vectors, and is
torsion free, \footnote{In this paper, we do not treat the
  torsion tensor. If we wish to consider a spacetime with torsion, we
  have to extend the formulation to that including the torsion tensor.}
we can regard it as a derivative operator on the background
spacetime ${\cal M}_{0}$\cite{Wald-book}. 
Of course, the representation of this derivative operator 
${\cal X}^{*}\bar{\nabla}_{a}\left({\cal X}^{-1}\right)^{*}$ on the
background spacetime ${\cal M}_{0}$ depends entirely on the gauge
choice ${\cal X}$.
Though we should keep in mind that we have already chosen a gauge
when we regard Eq.~(\ref{eq:physical-spacetime-Riemann-def}) as
an equation on the background spacetime ${\cal M}_{0}$, we do not
explicitly express the index of the gauge choice ${\cal X}$ in any
expression, again.


Since $\bar{\nabla}_{a}$ 
($={\cal X}^{*}\bar{\nabla}_{a}\left({\cal X}^{-1}\right)^{*}$) 
may be regarded as a derivative operator on the background
spacetime that satisfies $\bar{\nabla}_{a}\bar{g}_{bc}=0$, there
exists a tensor field $C^{c}_{\;\;ab}$ on the background
spacetime ${\cal M}_{0}$ such that 
\begin{equation}
  \bar{\nabla}_{a}\omega_{b}
  = \nabla_{a}\omega_{b} - C^{c}_{\;\;ab} \omega_{c},
\end{equation}
where $\omega_{c}$ is an arbitrary one-form on the background
spacetime ${\cal M}_{0}$.
From the property
(\ref{eq:property-as-covariant-derivative-on-phys-sp}) of the
covariant derivative operator $\bar{\nabla}_{a}$ on ${\cal M}$, the
tensor field $C^{c}_{\;\;ab}$ on ${\cal M}_{0}$ is given by
\begin{equation}
  C^{c}_{\;\;ab} = \frac{1}{2} \bar{g}^{cd}
  \left(
      \nabla_{a}\bar{g}_{db}
    + \nabla_{b}\bar{g}_{da}
    - \nabla_{d}\bar{g}_{ab}
  \right).
  \label{eq:c-connection}
\end{equation}
We note that the gauge dependence of the derivative
$\bar{\nabla}_{a}$ as an operator on ${\cal M}_{0}$ is included only
in this tensor field $C^{c}_{\;\;ab}$.
From Eq.~(\ref{eq:physical-spacetime-Riemann-def}), the Riemann
curvature $\bar{R}_{abc}^{\;\;\;\;\;\;d}$ associated with the metric
$\bar{g}_{ab}$ is given by the Riemann curvature
$R_{abc}^{\;\;\;\;\;\;d}$ on the background spacetime and the tensor
field $C^{c}_{\;\;ab}$ as follows:
\begin{equation}
  \bar{R}_{abc}^{\;\;\;\;\;\;d} = R_{abc}^{\;\;\;\;\;\;d}
  - 2 \nabla_{[a}^{} C^{d}_{\;\;b]c}
  + 2 C^{e}_{\;\;c[a} C^{d}_{\;\;b]e}.
  \label{eq:phys-riemann-back-riemann-rel}
\end{equation}
To obtain a perturbative expression of the curvatures, we first
calculate the expansion of the inverse metric $\bar{g}^{ab}$, and then
the perturbative expression of the tensor $C^{c}_{\;\;ab}$ by using
Eq.~(\ref{eq:c-connection}). 
Next, we derive an expression of the perturbative curvature.


In this paper, we present some formulae for the second-order
perturbative curvature in the two-parameter perturbation theory.
To derive the second-order formulae, we first calculate the
$O(\epsilon\lambda)$ formulae, since the other second-order
formulae [$O(\epsilon^{2})$ and $O(\lambda^{2})$] are easily
derived from these for $O(\epsilon\lambda)$ through a simple
replacement of the perturbative variables.
We also note that all variables on the physical spacetime ${\cal M}$
are pulled-back to the background spacetime ${\cal M}_{0}$ using a
gauge choice ${\cal X}$.
In this sense, all variables treated below are tensor fields
defined on the background spacetime ${\cal M}_{0}$.
We also denote the perturbative expansion of the pull-back of the
variable $\bar{Q}$ on the physical spacetime ${\cal M}$ by
\begin{eqnarray}
  \bar{Q}
  =
  \sum_{k',k=0}^{\infty}
  \frac{\epsilon^{k}\lambda^{k'}}{k!k'!}
  {}^{(k,k')}\!\bar{Q} , 
  \label{eq:expansion-form}
\end{eqnarray}
as in Eq.~(\ref{eq:two-parameter-perturbation-def}).


Once we have derived the formulae of the perturbative Riemann
curvature (see \S\ref{sec:inverse-metric}) of each order, it is
straightforward to derive corresponding formulae of the Ricci
curvature (\S\ref{sec:Ricci-tensor}), scalar curvature
(\S\ref{sec:scalar-curvature}), Einstein tensor
(\S\ref{sec:Einstein-tensor}), and Weyl curvature (\S\ref{sec:weyl}). 
We also derive the perturbative form of the divergence of an
arbitrary tensor field of the second rank to check the
perturbative Bianchi identities.


\subsection{Expansion of the inverse metric and the Riemann curvature}
\label{sec:inverse-metric}


Following the outline of the calculations explained above,
we first calculate the perturbative expansion of the inverse metric.
The expression for the inverse metric can be readily derived from the
expansion (\ref{eq:metric-expansion}) of the metric $\bar{g}_{ab}$ and
the definition of the inverse metric 
\begin{equation}
  \bar{g}^{ab}\bar{g}_{bc} = \delta^{a}_{\;\;c}.
\end{equation}
We also expand the inverse metric $\bar{g}^{ab}$ in the form
(\ref{eq:expansion-form}).
Then, each term of the expansion of the inverse metric is given by
\begin{eqnarray}
  \label{eq:inverse-metric-first-order}
  {}^{(p,q)}\!\bar{g}^{ab}
  &=& - {}^{(p,q)}\!h^{ab},
  \quad\quad\quad\quad\quad\quad\quad\quad\quad\quad
  (p,q) = (1,0), (0,1) \\
  \label{eq:inverse-metric-second-order}
  {}^{(p,q)}\!\bar{g}^{ab}
  &=&
  2 {}^{(\frac{p}{2},\frac{q}{2})}\!h^{ac}
    \; {}^{(\frac{p}{2},\frac{q}{2})}\!h_{c}^{\;\;b}
  - {}^{(p,q)}\!h^{ab},
  \quad\quad\quad
  (p,q) = (2,0), (0,2) \\
  \label{eq:inverse-metric-epsilonlambda}
  {}^{(1,1)}\!\bar{g}^{ab}
  &=&
  {}^{(0,1)}\!h^{ca} \; {}^{(1,0)}\!h_{c}^{\;\;b}
  + {}^{(0,1)}\!h^{cb} \; {}^{(1,0)}\!h_{c}^{\;\;a}
  - {}^{(1,1)}\!h^{ab}.
\end{eqnarray}


To derive the formulae for the perturbative expansion of the Riemann
curvature, we have to derive the formulae for the perturbative
expansion of the tensor $C^{c}_{\;\;ab}$ defined in
Eq.~(\ref{eq:c-connection}).
The tensor $C^{c}_{\;\;ab}$ is also expanded in the same form as
Eq.~(\ref{eq:expansion-form}).
The first-order perturbations of $C^{c}_{\;\;ab}$ have the well-known
form\cite{Wald-book} 
\begin{eqnarray}
  {}^{(p,q)}\!C^{c}_{\;\;ab}
  &=&
  \nabla_{(a}\;{}^{(p,q)}\!h_{b)}^{\;\;\;c}
  - \frac{1}{2} \nabla^{c}\;{}^{(p,q)}\!h_{ab}
  =:
  H_{ab}^{\;\;\;\;c}\left[{}^{(p,q)}\!h\right],
  \label{eq:(1)H-def}
\end{eqnarray}
where $(p,q)=(1,0),(0,1)$, and ${}^{(p,q)}\!h$ in the brackets of
the variable $H_{ab}^{\;\;\;\;c}\left[{}^{(p,q)}\!h\right]$
indicates that $H_{ab}^{\;\;\;\;c}\left[{}^{(p,q)}\!h\right]$ is
constituted of three covariant derivatives of the perturbative
metric ${}^{(p,q)}\!h_{ab}$.  
In terms of the tensor field defined by Eq.~(\ref{eq:(1)H-def}),
the second-order perturbations of $C^{c}_{\;\;ab}$ are given by
\begin{eqnarray}
  \label{eq:C-expand-second-1}
  {}^{(p,q)}\!C^{c}_{\;\;ab} &=&
  H^{\;\;\;\;c}_{ab}\left[{}^{(p,q)}\!h\right]
  - 2 {}^{(\frac{p}{2},\frac{q}{2})}\!h^{cd}
  \;
  H_{abd} 
  \left[{}^{(\frac{p}{2},\frac{q}{2})}\!h\right],
  \\
  {}^{(1,1)}\!C^{c}_{\;\;ab} &=&
  H^{\;\;\;\;c}_{ab}\left[{}^{(1,1)}\!h\right]
  - {}^{(1,0)}\!h^{cd}
  \;
  H_{abd}\left[{}^{(0,1)}\!h\right]
  - {}^{(0,1)}\!h^{cd}
  \;
  H_{abd}\left[{}^{(1,0)}\!h\right],
  \label{eq:C-expand-second-2}
\end{eqnarray}
where $(p,q) = (2,0), (0,2)$ in Eq.~(\ref{eq:C-expand-second-1}).


The Riemann curvature (\ref{eq:phys-riemann-back-riemann-rel})
on the physical spacetime ${\cal M}$ can be expanded in the form
(\ref{eq:expansion-form}). 
The forms of the perturbative Riemann curvature up to second
order are given by
\begin{eqnarray}
  \label{eq:Riemann-expand-first}
  {}^{(p,q)}\!\bar{R}_{abc}^{\;\;\;\;\;\;d}
  &=&
  - 2 \nabla_{[a}\;{}^{(p,q)}\!C^{d}_{\;\;b]c},
  \quad\quad\quad\quad\quad\quad\quad\quad\quad\quad
  (p,q)=(1,0),(0,1)
  , \\
  \label{eq:Riemann-expand-second-1}
  {}^{(p,q)}\!\bar{R}_{abc}^{\;\;\;\;\;\;d}
  &=&
  - 2 \nabla_{[a}\;{}^{(p,q)}\!C^{d}_{\;\;b]c}
  + 4 \;{}^{(\frac{p}{2},\frac{q}{2})}\!C^{e}_{\;\;c[a}
      \;{}^{(\frac{p}{2},\frac{q}{2})}\!C^{d}_{\;\;b]e},
  \;
  (p,q)=(2,0),(0,2)
  , \\
  \label{eq:Riemann-expand-second-2}
  {}^{(1,1)}\!\bar{R}_{abc}^{\;\;\;\;\;\;d}
  &=&
  - 2 \nabla_{[a}\;{}^{(1,1)}\!C^{d}_{\;\;b]c}
  + 2 \;{}^{(1,0)}\!C^{e}_{\;\;c[a}\;{}^{(0,1)}\!C^{d}_{\;\;b]e}
  + 2 \;{}^{(0,1)}\!C^{e}_{\;\;c[a}\;{}^{(0,1)}\!C^{d}_{\;\;b]e}.
\end{eqnarray}
Substituting Eqs.~(\ref{eq:(1)H-def})--(\ref{eq:C-expand-second-2}) into
Eqs.~(\ref{eq:Riemann-expand-first})--(\ref{eq:Riemann-expand-second-2}),
we obtain the perturbative form of the Riemann curvature in terms
of the variables defined by Eq.~(\ref{eq:(1)H-def}).
This perturbative form of linear-order is simply given by the
replacement 
\begin{equation}
  {}^{(p,q)}\!C^{c}_{\;\;ab}
  \rightarrow 
  H_{ab}^{\;\;\;\;c}\left[{}^{(p,q)}\!h\right]
\end{equation}
in Eq.~(\ref{eq:Riemann-expand-first}).
On the other hand, the perturbative form of the
$O(\epsilon^{2})$ and $O(\lambda^{2})$ Riemann curvatures are
derived from the perturbative form of $O(\epsilon\lambda)$.
For these reasons, we only present the derivation of the perturbative
form of $O(\epsilon\lambda)$ in terms of the variables defined by
Eq.~(\ref{eq:(1)H-def}),
\begin{eqnarray}
  {}^{(1,1)}\!\bar{R}_{abc}^{\;\;\;\;\;d}
  &=&
  - 2 \nabla_{[a}^{} H_{b]c}^{\;\;\;\;d}\left[{}^{(1,1)}\!h\right]
  \nonumber\\
  &&
  \quad
  + 2 H_{[a}^{\;\;\;de}\left[{}^{(1,0)}\!h\right]
      H_{b]ce}^{}\left[{}^{(0,1)}\!h\right]
  + 2 H_{[a}^{\;\;\;de}\left[{}^{(0,1)}\!h\right]
      H_{b]ce}^{}\left[{}^{(1,0)}\!h\right]
  \nonumber\\
  &&
  \quad
  + 2 {}^{(1,0)}\!h^{de}
      \nabla_{[a}^{}\;H_{b]ce}^{}\left[{}^{(0,1)}\!h\right]
  + 2 {}^{(0,1)}\!h^{de}
      \nabla_{[a}^{}\;H_{b]ce}^{}\left[{}^{(1,0)}\!h\right],
  \label{eq:1.1-pertrubative-curvatures-(1)H-def}
\end{eqnarray}
as the second-order perturbative curvature.


To write down the perturbative curvatures
(\ref{eq:Riemann-expand-first}) and
(\ref{eq:1.1-pertrubative-curvatures-(1)H-def}) in terms of the gauge 
invariant and variant variables defined by
Eqs.~(\ref{eq:linear-metric-decomp})--(\ref{eq:H-ab-in-gauge-X-def-second-2}),
we first derive an expression for the tensor field
$H_{ab}^{\;\;\;\;c}\left[{}^{(p,q)}\!h\right]$ in terms
of the gauge invariant variables, and then, we derive a perturbative
expression for the Riemann curvature.


First, we consider the linear-order perturbation
(\ref{eq:Riemann-expand-first}) of the Riemann curvature.
Using the decomposition (\ref{eq:linear-metric-decomp}) and the
identity $R_{[abc]}^{\;\;\;\;\;\;\;\;d} = 0$, we can easily
derive the relation
\begin{eqnarray}
  H_{abc}\left[{}^{(p,q)}\!h\right]
  =
  H_{abc}\left[{}^{(p,q)}\!{\cal H}\right]
  + \nabla_{a}\nabla_{b}\;{}^{(p,q)}\!X_{c}
  + R_{bca}^{\;\;\;\;\;\;d}\;{}^{(p,q)}\!X_{d},
  \label{eq:3.5}
\end{eqnarray}
where the variable 
$H_{abc}\left[{}^{(p,q)}\!{\cal H}\right]$ is defined by
\begin{eqnarray}
  H_{abc}\left[{}^{(p,q)}\!{\cal H}\right] 
  &:=& 
  g_{cd} H_{ab}^{\;\;\;\;d}\left[{}^{(p,q)}\!{\cal H}\right],
  \\
  \quad
  H_{ab}^{\;\;\;\;c}\left[{}^{(p,q)}\!{\cal H}\right]
  &:=&
  \nabla_{(a}\;{}^{(p,q)}\!{\cal H}_{b)}^{\;\;\;c}
  - \frac{1}{2} \nabla^{c}\;{}^{(p,q)}\!{\cal H}_{ab}.
  \label{eq:(1)calH-def}
\end{eqnarray}
Clearly, the variable 
$H_{ab}^{\;\;\;\;c}\left[{}^{(p,q)}\!{\cal H}\right]$
is gauge invariant.  
Taking the derivative of $H_{abc}$ and using the Bianchi
identity $\nabla_{[a}R_{bc]de}=0$, we obtain 
\begin{eqnarray}
  {}^{(p,q)}\!\bar{R}_{abc}^{\;\;\;\;\;\;d} &=&
  - 2 \nabla_{[a}^{}H_{b]c}^{\;\;\;\;\;d}
  \left[{}^{(p,q)}\!{\cal H}\right]
  + {\pounds}_{{}^{(p,q)}\!X}R_{abc}^{\;\;\;\;\;\;d},
  \label{eq:linear-order-perturbation-riemann}
\end{eqnarray}
where $(p,q)=(1,0),(0,1)$.


Next, we consider the second-order curvature perturbation.
We first consider the $O(\epsilon\lambda)$ metric perturbation
as mentioned above.
Inspecting the definition
(\ref{eq:H-ab-in-gauge-X-def-second-2}) of the gauge invariant
variable of $O(\epsilon\lambda)$ metric perturbation, we first
define the variable
\begin{eqnarray}
  {}^{(1,1)}\!\widehat{\cal H}_{ab}
  := {}^{(1,1)}\!h_{ab}
  - {\pounds}_{{}^{(0,1)}\!X} {}^{(1,0)}\!h_{ab}
  - {\pounds}_{{}^{(1,0)}\!X} {}^{(0,1)}\!h_{ab}
  \nonumber\\
  \quad\quad
  + \frac{1}{2} \left(
      {\pounds}_{{}^{(1,0)}\!X}
      {\pounds}_{{}^{(0,1)}\!X}
    + {\pounds}_{{}^{(0,1)}\!X}
      {\pounds}_{{}^{(1,0)}\!X}
  \right) g_{ab}.
  \label{eq:11-hat-calH-ab-def}
\end{eqnarray}
As in the case of linear order, we evaluate the tensor
$H_{ab}^{\;\;\;\;c}\left[{}^{(1,1)}\!\widehat{\cal H}\right]$
and obtain
\begin{eqnarray}
  2 H_{ab}^{\;\;\;\;c}\left[{}^{(1,1)}\!\widehat{\cal H}\right]
  &=&
  2 H_{ab}^{\;\;\;\;c}\left[{}^{(1,1)}\!h\right]
  \nonumber\\
  && 
  - {\pounds}_{{}^{(0,1)}\!X}
  \left(
      H_{ab}^{\;\;\;\;c}\left[{}^{(1,0)}\!h\right]
    + H_{ab}^{\;\;\;\;c}\left[{}^{(1,0)}\!{\cal H}\right]
  \right)
  \nonumber\\
  && 
  - {\pounds}_{{}^{(1,0)}\!X}
  \left(
      H_{ab}^{\;\;\;\;c}\left[{}^{(0,1)}\!h\right]
    + H_{ab}^{\;\;\;\;c}\left[{}^{(0,1)}\!{\cal H}\right]
  \right)
  \nonumber\\
  && 
  +
  \left(
      H_{abd}\left[{}^{(1,0)}\!h\right]
    + H_{abd}\left[{}^{(1,0)}\!{\cal H}\right]
  \right)
  {\pounds}_{{}^{(0,1)}\!X}g^{cd}
  \nonumber\\
  && 
  +
  \left(
      H_{ab}^{\;\;\;\;c}\left[{}^{(0,1)}\!h\right]
    + H_{ab}^{\;\;\;\;c}\left[{}^{(0,1)}\!{\cal H}\right]
  \right)
  {\pounds}_{{}^{(1,0)}\!X}g^{cd}
  \nonumber\\
  && 
  - \left(
    {}^{(1,0)}\!h_{d}^{\;\;e} + {}^{(1,0)}\!{\cal H}_{d}^{\;\;e}
  \right)
  \left(
    \nabla_{a}\nabla_{b}{}^{(0,1)}\!X^{d} 
    - R_{eab}^{\;\;\;\;\;\;d}{}^{(0,1)}\!X^{e}
  \right)
  \nonumber\\
  && 
  - \left(
    {}^{(0,1)}\!h_{d}^{\;\;e} + {}^{(0,1)}\!{\cal H}_{d}^{\;\;e}
  \right)
  \left(
    \nabla_{a}\nabla_{b}{}^{(1,0)}\!X^{d} 
    - R_{eab}^{\;\;\;\;\;\;d}{}^{(1,0)}\!X^{e}
  \right).
\end{eqnarray}
In the derivation of this expression, some formulae, which are
summarized in Appendix \ref{sec:use-ful}, are useful. 
After straightforward calculations, we obtain 
\begin{eqnarray}
  {}^{(1,1)}\!\bar{R}_{abc}^{\;\;\;\;\;\;d}
  &=&
  - 2 \nabla_{[a}^{}
  H_{b]c}^{\;\;\;\;\;d}\left[{}^{(1,1)}\!\widehat{\cal H}\right]
  \nonumber\\
  && 
  + 2 H_{[a}^{\;\;\;de}\left[{}^{(1,0)}\!{\cal H}\right]
      H_{b]ce}^{}\left[{}^{(0,1)}\!{\cal H}\right]
  + 2 H_{[a}^{\;\;\;de}\left[{}^{(0,1)}\!{\cal H}\right]
      H_{b]ce}^{}\left[{}^{(1,0)}\!{\cal H}\right]
  \nonumber\\
  && 
  + {}^{(1,0)}\!{\cal H}_{e}^{\;\;d} 
  \left(
      {\pounds}_{{}^{(0,1)}\!X} R_{abc}^{\;\;\;\;\;e}
    - {}^{(0,1)}\!\bar{R}_{abc}^{\;\;\;\;\;\;e}
  \right)
  \nonumber\\
  &&
  + {}^{(0,1)}\!{\cal H}_{e}^{\;\;d} 
  \left(
      {\pounds}_{{}^{(1,0)}\!X} R_{abc}^{\;\;\;\;\;e}
    - {}^{(1,0)}\!\bar{R}_{abc}^{\;\;\;\;\;\;e}
  \right)
  \nonumber\\
  && 
  + {\pounds}_{{}^{(0,1)}\!X}
  \left(
      {}^{(1,0)}\!\bar{R}_{abc}^{\;\;\;\;\;\;d}
    - \frac{1}{2} {\pounds}_{{}^{(1,0)}\!X} R_{abc}^{\;\;\;\;\;d}
  \right)
  \nonumber\\
  && 
  + {\pounds}_{{}^{(1,0)}\!X}
  \left(
      {}^{(0,1)}\!\bar{R}_{abc}^{\;\;\;\;\;\;d}
    - \frac{1}{2} {\pounds}_{{}^{(0,1)}\!X} R_{abc}^{\;\;\;\;\;d}
  \right).
\end{eqnarray} 
The first term on the right-hand side,
$-2 \nabla_{[a}^{}H_{b]c}^{\;\;\;\;\;d}\left[{}^{(1,1)}\!\widehat{\cal H}\right]$,
includes the gauge degree of freedom, because the variable
${}^{(1,1)}\!\widehat{\cal H}_{ab}$ is transformed as a linear-order metric
perturbation.
Since we have already assume that the linear metric perturbation is
decomposed as Eq.~(\ref{eq:linear-metric-decomp}), we can also 
decomposed the variable ${}^{(1,1)}\!\widehat{\cal H}_{ab}$
as
\begin{equation}
  {}^{(1,1)}\!\widehat{\cal H}_{ab} =: {}^{(1,1)}\!{\cal H}_{ab}
  + 2 \nabla_{(a} {}^{(1,1)}\!X_{b)}, 
\end{equation}
as pointed out in a previous paper\cite{kouchan-gauge-inv}.
The variables ${}^{(1,1)}\!{\cal H}_{ab}$ and ${}^{(1,1)}\!X_{b}$ are
gauge invariant and variant parts of $O(\epsilon\lambda)$ metric
perturbation.
Then, as in the case of linear order, we obtain
\begin{eqnarray}
  - 2 \nabla_{[a}^{}H_{b]c}^{\;\;\;\;\;d}\left[{}^{(1,1)}\!\widehat{\cal H}\right]
  &=& 
  - 2 \nabla_{[a}^{}H_{b]c}^{\;\;\;\;\;d}\left[{}^{(1,1)}\!{\cal H}\right]
  + {\pounds}_{{}^{(1,1)}\!X} R_{abc}^{\;\;\;\;\;\;d}.
\end{eqnarray}
Further, using Eq.~(\ref{eq:linear-order-perturbation-riemann}), we
reach the final form of the perturbative Riemann curvature of
$O(\epsilon\lambda)$:
\begin{eqnarray}
  {}^{(1,1)}\!\bar{R}_{abc}^{\;\;\;\;\;\;d}
  &=&
  - 2 \nabla_{[a}^{}
  H_{b]c}^{\;\;\;\;\;d}\left[{}^{(1,1)}\!{\cal H}\right]
  \nonumber\\
  && \quad
  + 2 H_{[a}^{\;\;\;de}\left[{}^{(1,0)}\!{\cal H}\right]
      H_{b]ce}^{}\left[{}^{(0,1)}\!{\cal H}\right]
  + 2 H_{[a}^{\;\;\;de}\left[{}^{(0,1)}\!{\cal H}\right]
      H_{b]ce}^{}\left[{}^{(1,0)}\!{\cal H}\right]
  \nonumber\\
  && \quad
  + 2 {}^{(1,0)}\!{\cal H}_{e}^{\;\;d} 
  \nabla_{[a}^{}
  H_{b]c}^{\;\;\;\;\;e}\left[{}^{(0,1)}\!{\cal H}\right]
  + 2 {}^{(0,1)}\!{\cal H}_{e}^{\;\;d} 
  \nabla_{[a}^{}
  H_{b]c}^{\;\;\;\;\;e}\left[{}^{(1,0)}\!{\cal H}\right]
  \nonumber\\
  && \quad
  + {\pounds}_{{}^{(0,1)}\!X} {}^{(1,0)}\!\bar{R}_{abc}^{\;\;\;\;\;\;d}
  + {\pounds}_{{}^{(1,0)}\!X} {}^{(0,1)}\!\bar{R}_{abc}^{\;\;\;\;\;\;d}
  \nonumber\\
  && \quad
  +
  \left(
                  {\pounds}_{{}^{(1,1)}\!X}
    - \frac{1}{2} {\pounds}_{{}^{(0,1)}\!X}
                  {\pounds}_{{}^{(1,0)}\!X}
    - \frac{1}{2} {\pounds}_{{}^{(1,0)}\!X}
                  {\pounds}_{{}^{(0,1)}\!X}
  \right) R_{abc}^{\;\;\;\;\;d}.
  \label{eq:epsilon-lambda-riemann-final}
\end{eqnarray} 
The first three lines on the right-hand side of 
Eq.~(\ref{eq:epsilon-lambda-riemann-final}) are the gauge
invariant part and the remaining two lines are the gauge variant
part of the perturbative Riemann curvature of $O(\epsilon\lambda)$.


The perturbative Riemann curvatures of $O(\epsilon^{2})$ and
$O(\lambda^{2})$ are simply obtained through the replacement of
the variables in Eq.~(\ref{eq:epsilon-lambda-riemann-final}) of
$O(\epsilon\lambda)$. 
To obtain the Riemann curvature of $O(\epsilon^{2})$, we
consider the replacements of the variables 
\begin{eqnarray}
  {}^{(0,1)}\!X^{a} \rightarrow {}^{(1,0)}\!X^{a}, \quad
  {}^{(0,1)}\!{\cal H}_{ab}
  \rightarrow
  {}^{(1,0)}\!{\cal H}_{ab}
  \label{eq:replace-1,1-to-2,0}
\end{eqnarray} 
in Eq.~(\ref{eq:epsilon-lambda-riemann-final}).
Similarly, to obtain the $O(\lambda^{2})$ Riemann curvature, 
we consider the replacements of the variables 
\begin{eqnarray}
  {}^{(1,0)}\!X^{a} \rightarrow {}^{(0,1)}\!X^{a}, \quad
  {}^{(1,0)}\!{\cal H}_{ab}
  \rightarrow
  {}^{(0,1)}\!{\cal H}_{ab}
  \label{eq:replace-1,1-to-0,2}
\end{eqnarray}
in Eq.~(\ref{eq:epsilon-lambda-riemann-final}).
These replacements are consistent with the definitions
(\ref{eq:H-ab-in-gauge-X-def-second-1}) and
(\ref{eq:H-ab-in-gauge-X-def-second-2}) of the gauge
invariant variables of $O(\epsilon^{2})$ and $O(\lambda^{2})$.
Hence, we obtain the perturbative forms of the Riemann
curvatures of $O(\epsilon^{2})$ and $O(\lambda^{2})$ as 
\begin{eqnarray}
  {}^{(p,q)}\!\bar{R}_{abc}^{\;\;\;\;\;\;d}
  &=&
  - 2 \nabla_{[a}^{} H_{b]c}^{\;\;\;\;\;d}\left[{}^{(p,q)}\!{\cal H}\right]
  \nonumber\\
  && \quad
  + 4 H_{[a}^{\;\;\;de}\left[{}^{(\frac{p}{2},\frac{p}{2})}\!{\cal H}\right]
      H_{b]ce}^{}\left[{}^{(\frac{p}{2},\frac{p}{2})}\!{\cal H}\right]
  + 4 {}^{(\frac{p}{2},\frac{p}{2})}\!{\cal H}_{e}^{\;\;d} 
  \nabla_{[a}^{}
  H_{b]c}^{\;\;\;\;\;e}\left[{}^{(\frac{p}{2},\frac{p}{2})}\!{\cal H}\right]
  \nonumber\\
  && \quad
  + 2 {\pounds}_{{}^{(\frac{p}{2},\frac{p}{2})}\!X} {}^{(\frac{p}{2},\frac{p}{2})}\!\bar{R}_{abc}^{\;\;\;\;\;\;d}
  +
  \left(
      {\pounds}_{{}^{(p,q)}\!X}
    - {\pounds}_{{}^{(\frac{p}{2},\frac{p}{2})}\!X}^{2}
  \right) R_{abc}^{\;\;\;\;\;d},
  \label{eq:epsilon2-or-lambda2-riemann-final}
\end{eqnarray}
where $(p,q)=(2,0),(0,2)$.


Equations (\ref{eq:linear-order-perturbation-riemann}),
(\ref{eq:epsilon-lambda-riemann-final}) and 
(\ref{eq:epsilon2-or-lambda2-riemann-final}) show that 
all variables defined by 
\begin{eqnarray}
  {}^{(p,q)}\!{\cal R}_{abc}^{\;\;\;\;\;\;d}
  :=
  {}^{(p,q)}\!\bar{R}_{abc}^{\;\;\;\;\;\;d}
  - {\pounds}_{{}^{(p,q)}\!X}R_{abc}^{\;\;\;\;\;\;d}
  \label{eq:linear-order-gauge-inv-riemann}
\end{eqnarray}
for $(p,q)=(1,0),(0,1)$,
\begin{eqnarray}
  {}^{(p,q)}\!{\cal R}_{abc}^{\;\;\;\;\;\;d}
  :=
  {}^{(p,q)}\!\bar{R}_{abc}^{\;\;\;\;\;\;d}
  - 2 {\pounds}_{{}^{(\frac{p}{2},\frac{p}{2})}\!X} {}^{(\frac{p}{2},\frac{p}{2})}\!\bar{R}_{abc}^{\;\;\;\;\;\;d}
  -
  \left(
      {\pounds}_{{}^{(p,q)}\!X}
    - {\pounds}_{{}^{(\frac{p}{2},\frac{p}{2})}\!X}^{2}
  \right) R_{abc}^{\;\;\;\;\;d}
  \label{eq:epsilon2-or-lambda2-gauge-inv-riemann}
\end{eqnarray}
for $(p,q)=(2,0),(0,2)$, and
\begin{eqnarray}
  {}^{(1,1)}\!{\cal R}_{abc}^{\;\;\;\;\;\;d}
  &:=&
  {}^{(1,1)}\!\bar{R}_{abc}^{\;\;\;\;\;\;d}
  - {\pounds}_{{}^{(0,1)}\!X} {}^{(1,0)}\!\bar{R}_{abc}^{\;\;\;\;\;\;d}
  - {\pounds}_{{}^{(1,0)}\!X} {}^{(0,1)}\!\bar{R}_{abc}^{\;\;\;\;\;\;d}
  \nonumber\\
  && \quad
  -
  \left(
                  {\pounds}_{{}^{(1,1)}\!X}
    - \frac{1}{2} {\pounds}_{{}^{(0,1)}\!X}
                  {\pounds}_{{}^{(1,0)}\!X}
    - \frac{1}{2} {\pounds}_{{}^{(1,0)}\!X}
                  {\pounds}_{{}^{(0,1)}\!X}
  \right) R_{abc}^{\;\;\;\;\;d}
  \label{eq:epsilon-lambda-gauge-inv-riemann}
\end{eqnarray}
are gauge invariant.
These indeed do have the same forms as the definitions
(\ref{eq:matter-gauge-inv-def-1.0})--(\ref{eq:matter-gauge-inv-def-1.1})
for each order gauge invariant variable for an arbitrary field.


Here, we derive the perturbative formulae of the Riemann curvature
$\bar{R}_{abcd}$, which are used in the derivation of the Weyl
curvature $\bar{C}_{abcd}$.
For this purpose, we first expand the definition 
\begin{equation}
  \bar{R}_{abcd} = \bar{g}_{ed} \bar{R}_{abc}^{\;\;\;\;\;\;e}.
\end{equation}
The form of the perturbation of $\bar{R}_{abcd}$ at each order
is derived from the formulae 
\begin{eqnarray}
  {}^{(p,q)}\!\bar{R}_{abcd} &=& 
  {}^{(p,q)}\!\bar{g}_{ed} R_{abc}^{\;\;\;\;\;\;e}
  + 
  g_{ed} {}^{(p,q)}\bar{R}_{abc}^{\;\;\;\;\;\;e},
  \quad
  (p,q) = (1,0),  (0,1), \\
  {}^{(1,1)}\!\bar{R}_{abcd} &=& 
  {}^{(1,1)}\!\bar{g}_{ed} R_{abc}^{\;\;\;\;\;\;e}
  +
  {}^{(1,0)}\!\bar{g}_{ed} {}^{(0,1)}\!\bar{R}_{abc}^{\;\;\;\;\;\;e}
  \nonumber\\
  && \quad
  +
  {}^{(0,1)}\!\bar{g}_{ed} {}^{(1,0)}\!\bar{R}_{abc}^{\;\;\;\;\;\;e}
  + 
  g_{ed} {}^{(1,1)}\!\bar{R}_{abc}^{\;\;\;\;\;\;e}.
\end{eqnarray}
The formulae for ${}^{(p,q)}\!\bar{R}_{abcd}$ with $(p,q)=(2,0),(0,2)$
are derived using the replacements (\ref{eq:replace-1,1-to-2,0}) and
(\ref{eq:replace-1,1-to-0,2}) of the perturbative variables as
in the case of the Riemann curvature $\bar{R}_{abc}^{\;\;\;\;\;d}$.
The explicit form of each order ${}^{(p,q)}\!\bar{R}_{abcd}$ is
as follows: 
\begin{eqnarray}
  \label{eq:Riemann-dddd-linear}
  {}^{(p,q)}\!\bar{R}_{abcd} &=& 
  - 2 \nabla_{[a}H_{b]cd}\left[
    {}^{(p,q)}\!{\cal H}
  \right]
  + {}^{(p,q)}\!{\cal H}_{d}^{\;\;e} R_{abce}
  + {\pounds}_{{}^{(p,q)}\!X}R_{abcd}
\end{eqnarray}
for $(p,q)=(1,0),(0,1)$,
\begin{eqnarray}
  \label{eq:Ricci-dddd-second}
  {}^{(p,q)}\!\bar{R}_{abcd}
  &=&
  - 2 \nabla_{[a}H_{b]cd}\left[
    {}^{(p,q)}\!{\cal H}
  \right]
  + R_{abce} {}^{(p,q)}\!{\cal H}_{d}^{\;\;e}
  + 4 H_{d[a}^{\;\;\;\;\;e}\left[
    {}^{(p,q)}\!{\cal H}
  \right]
      H_{b]ce}\left[
    {}^{(p,q)}\!{\cal H}
  \right]
  \nonumber\\
  && \quad
  + 2 {\pounds}_{{}^{(\frac{p}{2},\frac{q}{2})}\!X} 
  {}^{(\frac{p}{2},\frac{q}{2})}\!R_{abcd}
  + \left(
    {\pounds}_{{}^{(p,q)}\!X}
    - {\pounds}_{{}^{(\frac{p}{2},\frac{q}{2})}\!X}^{2}
  \right)
  R_{abcd}
\end{eqnarray}
for $(p,q)=(2,0),(0,2)$, and
\begin{eqnarray}
  \label{eq:Riemann-dddd-1,1}
  {}^{(1,1)}\!\bar{R}_{abcd}
  &=&
  - 2 \nabla_{[a}H_{b]cd}\left[
    {}^{(1,1)}\!{\cal H}
  \right]
  + R_{abce} {}^{(1,1)}\!{\cal H}_{d}^{\;\;e}
  \nonumber\\
  && \quad
  + 2 H_{d[a}^{\;\;\;\;\;e}\left[
    {}^{(1,0)}\!{\cal H}
  \right]
      H_{b]ce}\left[
    {}^{(0,1)}\!{\cal H}
  \right]
  + 2 H_{d[a}^{\;\;\;\;\;e}\left[
    {}^{(0,1)}\!{\cal H}
  \right]
      H_{b]ce}\left[
    {}^{(1,0)}\!{\cal H}
  \right]
  \nonumber\\
  && \quad
  + {\pounds}_{{}^{(1,0)}\!X} {}^{(0,1)}\!R_{abcd}
  + {\pounds}_{{}^{(0,1)}\!X} {}^{(1,0)}\!R_{abcd}
  \nonumber\\
  && \quad
  + \left(
    {\pounds}_{{}^{(1,1)}\!X}
    - \frac{1}{2}{\pounds}_{{}^{(1,0)}\!X}
                 {\pounds}_{{}^{(0,1)}\!X}
    - \frac{1}{2}{\pounds}_{{}^{(1,0)}\!X}
                 {\pounds}_{{}^{(0,1)}\!X}
  \right)
  R_{abcd}.
\end{eqnarray}
The perturbative forms
(\ref{eq:Riemann-dddd-linear})--(\ref{eq:Riemann-dddd-1,1}) also show
that the variables defined by 
\begin{eqnarray}
  {}^{(p,q)}\!{\cal R}_{abcd}
  &:=&
  {}^{(p,q)}\!\bar{R}_{abcd}
  - {\pounds}_{{}^{(p,q)}\!X}R_{abcd}
  , \quad\quad\quad\quad\quad  (p,q)=(1,0), (0,1),
  \label{eq:linear-order-gauge-inv-riemann-dddd}
  \\
  {}^{(p,q)}\!{\cal R}_{abcd}
  &:=&
  {}^{(p,q)}\!\bar{R}_{abcd}
  - 2 {\pounds}_{{}^{(\frac{p}{2},\frac{p}{2})}\!X}
  {}^{(\frac{p}{2},\frac{p}{2})}\!\bar{R}_{abcd}
  \nonumber\\
  && \quad
  -
  \left(
      {\pounds}_{{}^{(p,q)}\!X}
    - {\pounds}_{{}^{(\frac{p}{2},\frac{p}{2})}\!X}^{2}
  \right) R_{abcd}
  , \quad\quad  (p,q)=(2,0), (0,2),
  \label{eq:epsilon2-or-lambda2-gauge-inv-riemann-dddd}
  \\
  {}^{(1,1)}\!{\cal R}_{abcd}
  &:=&
  {}^{(1,1)}\!\bar{R}_{abcd}
  - {\pounds}_{{}^{(0,1)}\!X} {}^{(1,0)}\!\bar{R}_{abcd}
  - {\pounds}_{{}^{(1,0)}\!X} {}^{(0,1)}\!\bar{R}_{abcd}
  \nonumber\\
  && \quad
  -
  \left(
                  {\pounds}_{{}^{(1,1)}\!X}
    - \frac{1}{2} {\pounds}_{{}^{(0,1)}\!X}
                  {\pounds}_{{}^{(1,0)}\!X}
    - \frac{1}{2} {\pounds}_{{}^{(1,0)}\!X}
                  {\pounds}_{{}^{(0,1)}\!X}
  \right) R_{abcd}
  \label{eq:epsilon-lambda-gauge-inv-riemann-dddd}
\end{eqnarray}
are gauge invariant.


\subsection{Ricci curvature}
\label{sec:Ricci-tensor}


Contracting the indices $b$ and $d$ in
Eqs.~(\ref{eq:linear-order-perturbation-riemann}),
(\ref{eq:epsilon-lambda-riemann-final}) and
(\ref{eq:epsilon2-or-lambda2-riemann-final}) of the perturbative
Riemann curvature, we can derive the formulae for the expansion of
the Ricci curvature,
\begin{eqnarray}
  {}^{(p,q)}\!\bar{R}_{ab} &=&
  - 2 \nabla_{[a}^{}H_{c]b}^{\;\;\;\;\;c}
  \left[{}^{(p,q)}\!{\cal H}\right]
  + {\pounds}_{{}^{(p,q)}\!X}R_{ab}
  \label{eq:linear-order-perturbation-ricci}
\end{eqnarray}
for first order and 
\begin{eqnarray}
  {}^{(p,q)}\!\bar{R}_{ab}
  &=&
  - 2 \nabla_{[a}^{}
  H_{c]b}^{\;\;\;\;\;c}\left[{}^{(p,q)}\!{\cal H}\right]
  + 4 H_{[a}^{\;\;\;cd}\left[{}^{(\frac{p}{2},\frac{p}{2})}\!{\cal H}\right]
      H_{c]bd}^{}\left[{}^{(\frac{p}{2},\frac{p}{2})}\!{\cal H}\right]
  \nonumber\\
  && \quad
  + 4 {}^{(\frac{p}{2},\frac{p}{2})}\!{\cal H}_{d}^{\;\;c} 
  \nabla_{[a}^{}
  H_{b]c}^{\;\;\;\;\;d}\left[{}^{(\frac{p}{2},\frac{p}{2})}\!{\cal H}\right]
  \nonumber\\
  && \quad
  + 2 {\pounds}_{{}^{(\frac{p}{2},\frac{p}{2})}\!X} 
      {}^{(\frac{p}{2},\frac{p}{2})}\!\bar{R}_{ab}
  +
  \left(
      {\pounds}_{{}^{(p,q)}\!X}
    - {\pounds}_{{}^{(\frac{p}{2},\frac{p}{2})}\!X}^{2}
  \right) R_{ab}
  \label{eq:epsilon2-or-lambda2-ricci}
  , \\
  {}^{(1,1)}\!\bar{R}_{ab}
  &=&
  - 2 \nabla_{[a}^{}
  H_{c]b}^{\;\;\;\;\;c}\left[{}^{(1,1)}\!{\cal H}\right]
  \nonumber\\
  && \quad
  + 2 H_{[a}^{\;\;\;cd}\left[{}^{(1,0)}\!{\cal H}\right]
      H_{c]bd}^{}\left[{}^{(0,1)}\!{\cal H}\right]
  + 2 H_{[a}^{\;\;\;cd}\left[{}^{(0,1)}\!{\cal H}\right]
      H_{c]bd}^{}\left[{}^{(1,0)}\!{\cal H}\right]
  \nonumber\\
  && \quad
  + 2 {}^{(1,0)}\!{\cal H}_{d}^{\;\;c} 
  \nabla_{[a}^{}
  H_{c]b}^{\;\;\;\;\;d}\left[{}^{(0,1)}\!{\cal H}\right]
  + 2 {}^{(0,1)}\!{\cal H}_{d}^{\;\;c} 
  \nabla_{[a}^{}
  H_{c]b}^{\;\;\;\;\;d}\left[{}^{(1,0)}\!{\cal H}\right]
  \nonumber\\
  && \quad
  + {\pounds}_{{}^{(0,1)}\!X} {}^{(1,0)}\!\bar{R}_{ab}
  + {\pounds}_{{}^{(1,0)}\!X} {}^{(0,1)}\!\bar{R}_{ab}
  \nonumber\\
  && \quad
  +
  \left(
                  {\pounds}_{{}^{(1,1)}\!X}
    - \frac{1}{2} {\pounds}_{{}^{(0,1)}\!X}
                  {\pounds}_{{}^{(1,0)}\!X}
    - \frac{1}{2} {\pounds}_{{}^{(1,0)}\!X}
                  {\pounds}_{{}^{(0,1)}\!X}
  \right) R_{ab},
  \label{eq:epsilon-lambda-ricci}
\end{eqnarray}
for second order, where $(p,q)=(1,0),(0,1)$ in
Eq.~(\ref{eq:linear-order-perturbation-ricci}) and
$(p,q)=(2,0),(0,2)$ in Eq.~(\ref{eq:epsilon2-or-lambda2-ricci}).


It is trivial from the derivation that 
Eqs. (\ref{eq:linear-order-perturbation-ricci}),
(\ref{eq:epsilon2-or-lambda2-ricci}) and 
(\ref{eq:epsilon-lambda-ricci}) show that the variables
defined by 
\begin{eqnarray}
  {}^{(p,q)}\!{\cal R}_{ab} &=&
  {}^{(p,q)}\!\bar{R}_{ab} - {\pounds}_{{}^{(p,q)}\!X}R_{ab}
  \label{eq:linear-order-gauge-inv-ricci}
\end{eqnarray}
for $(p,q)=(1,0),(0,1)$,
\begin{eqnarray}
  {}^{(p,q)}\!{\cal R}_{ab}
  &=&
  {}^{(p,q)}\!\bar{R}_{ab}
  - 2 {\pounds}_{{}^{(\frac{p}{2},\frac{p}{2})}\!X} {}^{(\frac{p}{2},\frac{p}{2})}\!\bar{R}_{ab}
  -
  \left(
      {\pounds}_{{}^{(p,q)}\!X}
    - {\pounds}_{{}^{(\frac{p}{2},\frac{p}{2})}\!X}^{2}
  \right) R_{ab}
  \label{eq:epsilon2-or-lambda2-gauge-inv-ricci}
\end{eqnarray}
for $(p,q)=(2,0),(0,2)$, and
\begin{eqnarray}
  {}^{(1,1)}\!{\cal R}_{ab}
  &=&
  {}^{(1,1)}\!\bar{R}_{ab}
  - {\pounds}_{{}^{(0,1)}\!X} {}^{(1,0)}\!\bar{R}_{ab}
  - {\pounds}_{{}^{(1,0)}\!X} {}^{(0,1)}\!\bar{R}_{ab}
  \nonumber\\
  && \quad
  -
  \left(
                  {\pounds}_{{}^{(1,1)}\!X}
    - \frac{1}{2} {\pounds}_{{}^{(0,1)}\!X}
                  {\pounds}_{{}^{(1,0)}\!X}
    - \frac{1}{2} {\pounds}_{{}^{(1,0)}\!X}
                  {\pounds}_{{}^{(0,1)}\!X}
  \right) R_{ab}
  \label{eq:epsilon-lambda-gauge-inv-ricci}
\end{eqnarray}
are gauge invariant. 
These also have the same forms as those in the definitions
(\ref{eq:matter-gauge-inv-def-1.0})--(\ref{eq:matter-gauge-inv-def-1.1})
of each order gauge invariant variable for the perturbation of an
arbitrary field.


\subsection{Scalar curvature}
\label{sec:scalar-curvature}


The scalar curvature on the physical spacetime ${\cal M}$ is
given by 
\begin{equation}
  \bar{R} := \bar{g}^{ab} \bar{R}_{ab}.
  \label{eq:scalar-curvature-def}
\end{equation}
To obtain the perturbative form of the scalar curvature, the
left-hand side of Eq.~(\ref{eq:scalar-curvature-def}) is
expanded in the form (\ref{eq:expansion-form}) and the
right-hand side is expanded by using the Leibniz rule.
Then, the perturbative formula for the scalar curvature at each
order is derived from perturbative form of the inverse metric 
(\ref{eq:inverse-metric-first-order})--(\ref{eq:inverse-metric-epsilonlambda})
and the Ricci curvature
(\ref{eq:linear-order-perturbation-ricci})--(\ref{eq:epsilon-lambda-ricci}).
Straightforward calculations lead to the expansion of the scalar
curvature.


Using (\ref{eq:inverse-metric-first-order}) and
(\ref{eq:linear-order-perturbation-ricci}), we obtain the
first-order perturbative form of the scalar curvature as
\begin{eqnarray}
  {}^{(p,q)}\!\bar{R} =
  - 2 \nabla_{[a}^{} H_{b]}^{\;\;\;ab}
  \left[{}^{(p,q)}\!{\cal H}\right]
  -   R_{ab} {}^{(p,q)}\!{\cal H}^{ab}
  +   {\pounds}_{{}^{(p,q)}\!X} R ,
  \label{eq:linear-order-perturbation-scalar}
\end{eqnarray}
where $(p,q)=(0,1),(1,0)$.
Then, using (\ref{eq:inverse-metric-first-order}),
(\ref{eq:inverse-metric-epsilonlambda}),
(\ref{eq:linear-order-perturbation-ricci}) and
(\ref{eq:epsilon-lambda-ricci}), the perturbative scalar
curvature of $O(\epsilon\lambda)$ is found to be given by 
\begin{eqnarray}
  {}^{(1,1)}\!\bar{R}
  &=& 
  - 2 \nabla_{[a}^{}H_{b]}^{\;\;\;ab}
  \left[{}^{(1,1)}\!{\cal H}\right]
  + R^{ab} \left(
    2 {}^{(1,0)}\!{\cal H}_{c(a}^{} {}^{(0,1)}\!{\cal H}_{b)}^{\;\;c} 
    - {}^{(1,1)}\!{\cal H}_{ab}
  \right)
  \nonumber\\
  && \quad
  + 2 H_{[a}^{\;\;\;cd}\left[{}^{(0,1)}\!{\cal H}\right]
      H_{c]\;\;d}^{\;\;\;a}\left[{}^{(1,0)}\!{\cal H}\right]
  + 2 H_{[a}^{\;\;\;cd}\left[{}^{(1,0)}\!{\cal H}\right]
      H_{c]\;\;d}^{\;\;\;a}\left[{}^{(0,1)}\!{\cal H}\right]
  \nonumber\\
  && \quad
  + 2 {}^{(1,0)}\!{\cal H}_{c}^{\;\;b}
  \nabla_{[a}^{}H_{b]}^{\;\;\;ac}\left[{}^{(0,1)}\!{\cal H}\right]
  + 2 {}^{(0,1)}\!{\cal H}_{c}^{\;\;b}
  \nabla_{[a}^{}H_{b]}^{\;\;\;ac}\left[{}^{(1,0)}\!{\cal H}\right]
  \nonumber\\
  && \quad
  + 2 {}^{(1,0)}\!{\cal H}^{ab}
  \nabla_{[a}^{}H_{d]b}^{\;\;\;\;\;d}\left[{}^{(0,1)}\!{\cal H}\right]
  + 2 {}^{(0,1)}\!{\cal H}^{ab}
  \nabla_{[a}^{}H_{d]b}^{\;\;\;\;\;d}\left[{}^{(1,0)}\!{\cal H}\right]
  \nonumber\\
  && \quad
  + {\pounds}_{{}^{(0,1)}\!X} {}^{(1,0)}\!\bar{R}
  + {\pounds}_{{}^{(1,0)}\!X} {}^{(0,1)}\!\bar{R}
  \nonumber\\
  && \quad
  +
  \left(
                  {\pounds}_{{}^{(1,1)}\!X}
    - \frac{1}{2} {\pounds}_{{}^{(0,1)}\!X}
                  {\pounds}_{{}^{(1,0)}\!X}
    - \frac{1}{2} {\pounds}_{{}^{(1,0)}\!X}
                  {\pounds}_{{}^{(0,1)}\!X}
  \right) R.
  \label{eq:epsilon-lambda-scalar}
\end{eqnarray}


To derive the perturbative scalar curvature of
$O(\epsilon^{2})$, the replacement (\ref{eq:replace-1,1-to-2,0})
of the variables is applied to Eq.~(\ref{eq:epsilon-lambda-scalar}). 
Similarly, the replacement (\ref{eq:replace-1,1-to-0,2})
is applied to Eq.~(\ref{eq:epsilon-lambda-scalar}) when we
derive the perturbative scalar curvature of $O(\lambda^{2})$. 
Then we obtain the perturbative form of the scalar curvatures of
$O(\epsilon^{2})$ and $O(\lambda^{2})$:
\begin{eqnarray}
  {}^{(p,q)}\!\bar{R}
  &=& 
  - 2 \nabla_{[a}^{}H_{b]}^{\;\;\;ab}\left[{}^{(p,q)}\!{\cal H}\right]
  + R^{ab} \left(
    2 {}^{(\frac{p}{2},\frac{p}{2})}\!{\cal H}_{ca}^{}
      {}^{(\frac{p}{2},\frac{p}{2})}\!{\cal H}_{b}^{\;\;c} 
    - {}^{(p,q)}\!{\cal H}_{ab}
  \right)
  \nonumber\\
  && \quad
  + 4 H_{[a}^{\;\;\;cd}\left[{}^{(\frac{p}{2},\frac{p}{2})}\!{\cal H}\right]
     H_{c]\;\;d}^{\;\;\;a}\left[{}^{(\frac{p}{2},\frac{p}{2})}\!{\cal H}\right]
  + 4 {}^{(\frac{p}{2},\frac{p}{2})}\!{\cal H}_{c}^{\;\;b}
  \nabla_{[a}^{}
  H_{b]}^{\;\;\;ac}\left[{}^{(\frac{p}{2},\frac{p}{2})}\!{\cal H}\right]
  \nonumber\\
  && \quad
  + 4 {}^{(\frac{p}{2},\frac{p}{2})}\!{\cal H}^{ab}
  \nabla_{[a}^{}
  H_{d]b}^{\;\;\;\;\;d}\left[{}^{(\frac{p}{2},\frac{p}{2})}\!{\cal H}\right]
  \nonumber\\
  && \quad
  + 2 {\pounds}_{{}^{(\frac{p}{2},\frac{p}{2})}\!X}
      {}^{(\frac{p}{2},\frac{p}{2})}\!\bar{R}
  +
  \left(
      {\pounds}_{{}^{(p,q)}\!X}
    - {\pounds}_{{}^{(\frac{p}{2},\frac{p}{2})}\!X}^{2}
  \right) R,
  \label{eq:epsilon2-lambda2-scalar}
\end{eqnarray}
where $(p,q)=(2,0),(0,2)$.


It is also trivial from the derivation that
Eqs.~(\ref{eq:linear-order-perturbation-scalar})--(\ref{eq:epsilon2-lambda2-scalar})
show that the variables defined by
\begin{eqnarray}
  {}^{(p,q)}\!{\cal R} &=&
  {}^{(p,q)}\!\bar{R} - {\pounds}_{{}^{(p,q)}\!X}R
  \label{eq:linear-order-gauge-inv-scalar}
\end{eqnarray}
for $(p,q)=(1,0),(0,1)$,
\begin{eqnarray}
  {}^{(p,q)}\!{\cal R}
  &=&
  {}^{(p,q)}\!\bar{R}
  - 2 {\pounds}_{{}^{(\frac{p}{2},\frac{p}{2})}\!X} {}^{(\frac{p}{2},\frac{p}{2})}\!\bar{R}
  -
  \left(
      {\pounds}_{{}^{(p,q)}\!X}
    - {\pounds}_{{}^{(\frac{p}{2},\frac{p}{2})}\!X}^{2}
  \right) R
  \label{eq:epsilon2-or-lambda2-gauge-inv-scalar}
\end{eqnarray}
for $(p,q)=(2,0),(0,2)$, and
\begin{eqnarray}
  {}^{(1,1)}\!{\cal R}
  &=&
  {}^{(1,1)}\!\bar{R}
  - {\pounds}_{{}^{(0,1)}\!X} {}^{(1,0)}\!\bar{R}
  - {\pounds}_{{}^{(1,0)}\!X} {}^{(0,1)}\!\bar{R}
  \nonumber\\
  && \quad
  -
  \left(
                  {\pounds}_{{}^{(1,1)}\!X}
    - \frac{1}{2} {\pounds}_{{}^{(0,1)}\!X}
                  {\pounds}_{{}^{(1,0)}\!X}
    - \frac{1}{2} {\pounds}_{{}^{(1,0)}\!X}
                  {\pounds}_{{}^{(0,1)}\!X}
  \right) R
  \label{eq:epsilon-lambda-gauge-inv-scalar}
\end{eqnarray}
are gauge invariant.
These too have the same forms as those given in the definitions
(\ref{eq:matter-gauge-inv-def-1.0})--(\ref{eq:matter-gauge-inv-def-1.1})
of each order gauge invariant variable for the perturbations of an
arbitrary field.


\subsection{Einstein tensor}
\label{sec:Einstein-tensor}


Next, we consider the perturbative form of the Einstein tensor.
The Einstein tensor on the physical spacetime ${\cal M}$ is
defined by
\begin{equation}
  \bar{G}_{ab}:=\bar{R}_{ab} - \frac{1}{2} \bar{g}_{ab}\bar{R}.
  \label{eq:Einstein-tensor-def}
\end{equation}
As in the case of the scalar curvature, the left-hand side of
Eq.~(\ref{eq:Einstein-tensor-def}) is expanded in the form
(\ref{eq:expansion-form}), and the second term on right-hand side
of Eq.~(\ref{eq:Einstein-tensor-def}) is expanded by using the
Leibniz rule.
Then, the perturbative formula for the Einstein tensor at each
order is derived from the perturbative form of the metric
(\ref{eq:linear-metric-decomp})--(\ref{eq:H-ab-in-gauge-X-def-second-2}),
those of the Ricci curvature
(\ref{eq:linear-order-perturbation-ricci})--(\ref{eq:epsilon-lambda-ricci}),
and those of the Ricci scalar
(\ref{eq:linear-order-perturbation-scalar})--(\ref{eq:epsilon2-lambda2-scalar}).


The linear order Einstein tensor is given by
\begin{eqnarray}
  {}^{(p,q)}\!\bar{G}_{ab} &=&
  - 2 \nabla_{[a}^{}
  {\cal H}_{d]b}^{\;\;\;\;\;d}\left[{}^{(p,q)}\!{\cal H}\right]
  + g_{ab}
  \nabla_{[c}^{}
  {\cal H}_{d]}^{\;\;\;cd}\left[{}^{(p,q)}\!{\cal H}\right]
  - \frac{1}{2} R \;{}^{(p,q)}\!{\cal H}_{ab}
  + \frac{1}{2} g_{ab} R_{cd} \;{}^{(p,q)}\!{\cal H}^{cd}
  \nonumber\\
  &&\quad
  + {\pounds}_{{}^{(p,q)}\!X} G_{ab},
  \label{eq:linear-order-perturbation-Einstein}
\end{eqnarray}
where $(p,q)=(1,0),(0,1)$.
Next, using
Eqs.~(\ref{eq:linear-metric-decomp})--(\ref{eq:H-ab-in-gauge-X-def-second-2}), 
(\ref{eq:epsilon-lambda-ricci}), (\ref{eq:linear-order-perturbation-scalar})
and (\ref{eq:epsilon-lambda-scalar}), the Einstein tensor of
$O(\epsilon\lambda)$ is found to be given by 
\begin{eqnarray}
  {}^{(1,1)}\!\bar{G}_{ab}
  &=&
  - 2 \nabla_{[a}^{}
  H_{c]b}^{\;\;\;\;c}\left[{}^{(1,1)}\!{\cal H}\right]
  \nonumber\\
  &&
  + 2 H_{[a}^{\;\;\;de}\left[{}^{(1,0)}\!{\cal H}\right]
      H_{d]be}\left[{}^{(0,1)}\!{\cal H}\right]
  + 2 H_{[a}^{\;\;\;de}\left[{}^{(0,1)}\!{\cal H}\right]
      H_{d]be}\left[{}^{(1,0)}\!{\cal H}\right]
  \nonumber\\
  &&
  + 2 {}^{(1,0)}\!{\cal H}_{e}^{\;\;d}
  \nabla_{[a}
  H_{d]b}^{\;\;\;\;\;e}\left[{}^{(0,1)}\!{\cal H}\right]
  + 2 {}^{(0,1)}\!{\cal H}_{e}^{\;\;d}
  \nabla_{[a}
  H_{d]b}^{\;\;\;\;\;e}\left[{}^{(1,0)}\!{\cal H}\right]
  \nonumber\\
  &&
  - \frac{1}{2}g_{ab}
  \left(
    - 2 \nabla_{[c}^{}
    H_{d]}^{\;\;\;cd}\left[{}^{(1,1)}\!{\cal H}\right]
    + R_{de} \left(
        2 {}^{(0,1)}\!{\cal H}_{c}^{\;\;d} \; {}^{(1,0)}\!{\cal H}^{ec}
      - {}^{(1,1)}\!{\cal H}^{de}
    \right)
  \right.
  \nonumber\\
  && \quad\quad\quad\quad
  \left.
    + 2 
    H_{[c}^{\;\;\;de}\left[{}^{(1,0)}\!{\cal H}\right]
    H_{d]\;\;e}^{\;\;\;c}\left[{}^{(0,1)}\!{\cal H}\right]
    + 2 
    H_{[c}^{\;\;\;de}\left[{}^{(0,1)}\!{\cal H}\right]
    H_{d]\;\;e}^{\;\;\;c}\left[{}^{(1,0)}\!{\cal H}\right]
  \right.
  \nonumber\\
  && \quad\quad\quad\quad
  \left.
    + 2 {}^{(1,0)}\!{\cal H}_{e}^{\;\;d}
    \nabla_{[c}^{}
    H_{d]}^{\;\;\;ce}\left[{}^{(0,1)}\!{\cal H}\right]
    + 2 {}^{(0,1)}\!{\cal H}_{e}^{\;\;d}
    \nabla_{[c}^{}
    H_{d]}^{\;\;\;ce}\left[{}^{(1,0)}\!{\cal H}\right]
  \right.
  \nonumber\\
  && \quad\quad\quad\quad
  \left.
    + 2 {}^{(1,0)}\!{\cal H}^{ce}
    \nabla_{[c}^{}
    H_{d]e}^{\;\;\;\;\;d}\left[{}^{(0,1)}\!{\cal H}\right]
    + 2 {}^{(0,1)}\!{\cal H}^{ce}
    \nabla_{[c}^{}
    H_{d]e}^{\;\;\;\;\;d}\left[{}^{(1,0)}\!{\cal H}\right]
  \right)
  \nonumber\\
  && \quad
  + {}^{(0,1)}\!{\cal H}_{ab}
  \left(
    \nabla_{[c}^{}
    H_{d]}^{\;\;\;cd}\left[{}^{(1,0)}\!{\cal H}\right]
    + \frac{1}{2} R_{cd} {}^{(0,1)}\!{\cal H}^{cd}
  \right)
  \nonumber\\
  && \quad
  + {}^{(1,0)}\!{\cal H}_{ab}
  \left(
    \nabla_{[c}^{}
    H_{d]}^{\;\;\;cd}\left[{}^{(0,1)}\!{\cal H}\right]
    + \frac{1}{2} R_{cd} {}^{(1,0)}\!{\cal H}^{cd}
  \right)
  \nonumber\\
  && \quad
  - \frac{1}{2} R \; {}^{(1,1)}\!{\cal H}_{ab}
  \nonumber\\
  && \quad
  + {\pounds}_{{}^{(0,1)}\!X} {}^{(1,0)}\!\bar{G}_{ab}
  + {\pounds}_{{}^{(1,0)}\!X} {}^{(0,1)}\!\bar{G}_{ab}
  \nonumber\\
  && \quad
  + \left(
    {\pounds}_{{}^{(1,1)}\!X}
    - \frac{1}{2} {\pounds}_{{}^{(0,1)}\!X} {\pounds}_{{}^{(1,0)}\!X}
    - \frac{1}{2} {\pounds}_{{}^{(1,0)}\!X} {\pounds}_{{}^{(0,1)}\!X}
  \right) G_{ab}.
  \label{eq:epsilon-lambda-Einstein}
\end{eqnarray}
Then, through the replacements (\ref{eq:replace-1,1-to-2,0}) and
(\ref{eq:replace-1,1-to-0,2}), the perturbative forms of the Einstein
tensor of $O(\epsilon^{2})$ and $O(\lambda^{2})$ are given by 
\begin{eqnarray}
  {}^{(p,q)}\!\bar{G}_{ab}
  &=&
  - 2 \nabla_{[a}
  H_{c]b}^{\;\;\;\;c}\left[{}^{(p,q)}\!{\cal H}\right]
  \nonumber\\
  && \quad
  + 4 
  H_{[a}^{\;\;\;cd}\left[{}^{(\frac{p}{2},\frac{q}{2})}\!{\cal H}\right]
  H_{c]bd}\left[{}^{(\frac{p}{2},\frac{q}{2})}\!{\cal H}\right]
  + 4 
  {}^{(p,q)}\!{\cal H}_{c}^{\;\;d} 
  \nabla_{[a} H_{d]b}^{\;\;\;\;\;c}
  \left[{}^{(\frac{p}{2},\frac{q}{2})}\!{\cal H}\right]
  \nonumber\\
  && \quad
  - \frac{1}{2}g_{ab}
  \left(
    - 2 \nabla_{[c}H_{d]}^{\;\;\;cd}\left[{}^{(p,q)}\!{\cal H}\right]
    + R_{de} \left(
      2 {}^{(\frac{p}{2},\frac{q}{2})}\!{\cal H}_{c}^{\;\;d} \;
        {}^{(\frac{p}{2},\frac{q}{2})}\!{\cal H}^{ec}
      - {}^{(p,q)}\!{\cal H}^{de}
    \right)
  \right.
  \nonumber\\
  && \quad\quad\quad\quad\quad
  \left.
    + 4 
    H_{[c}^{\;\;\;de}\left[{}^{(\frac{p}{2},\frac{q}{2})}\!{\cal H}\right]
    H_{d]\;\;e}^{\;\;\;c}\left[{}^{(\frac{p}{2},\frac{q}{2})}\!{\cal H}\right]
  \right.
  \nonumber\\
  && \quad\quad\quad\quad\quad
  \left.
    + 4
    {}^{(\frac{p}{2},\frac{q}{2})}\!{\cal H}_{e}^{\;\;d} 
    \nabla_{[c}^{}H_{d]}^{\;\;\;ce}
    \left[{}^{(\frac{p}{2},\frac{q}{2})}\!{\cal H}\right]
    + 4
    {}^{(\frac{p}{2},\frac{q}{2})}\!{\cal H}^{ce} 
    \nabla_{[c}^{}H_{d]e}^{\;\;\;\;\;d}
    \left[{}^{(\frac{p}{2},\frac{q}{2})}\!{\cal H}\right]
  \right)
  \nonumber\\
  && \quad
  + 2 
  {}^{(\frac{p}{2},\frac{q}{2})}\!{\cal H}_{ab}
  \nabla_{[c}H_{d]}^{\;\;\;cd}
  \left[{}^{(\frac{p}{2},\frac{q}{2})}\!{\cal H}\right]
  + 
  {}^{(\frac{p}{2},\frac{q}{2})}\!{\cal H}_{ab}
  R_{cd}
  {}^{(\frac{p}{2},\frac{q}{2})}\!{\cal H}^{cd}
  - \frac{1}{2} 
  R 
  \;{}^{(p,q)}\!{\cal H}_{ab}
  \nonumber\\
  && \quad
  + 2 {\pounds}_{{}^{(\frac{p}{2},\frac{q}{2})}\!X} 
      {}^{(\frac{p}{2},\frac{q}{2})}\!\bar{G}_{ab}
  + 
  \left(
      {\pounds}_{{}^{(p,q)}\!X}
    - {\pounds}_{{}^{(\frac{p}{2},\frac{q}{2})}\!X}^{2}
  \right) G_{ab},
  \label{eq:epsilon2-lambda2-Einstein-dd}
\end{eqnarray}
where $(p,q)=(2,0),(0,2)$.


Further, we also derive the formulae for the perturbation of the
Einstein tensor 
\begin{equation}
  \bar{G}_{a}^{\;\;b} := \bar{g}^{bc} \bar{G}_{ac}
  \label{eq:Einstein-tensor-def-up-down}
\end{equation}
on ${\cal M}$.
Because the derivation of these formulae is similar to the above
perturbative curvatures, we only present the final results:
\begin{eqnarray}
  \label{eq:linear-Einstein}
  {}^{(p,q)}\!\bar{G}_{a}^{\;\;b}
  =
  {}^{(1)}{\cal G}_{a}^{\;\;b}\left[{}^{(p,q)}\!{\cal H}\right]
  + {\pounds}_{\displaystyle {}^{(p,q)}_{}X} \;\; G_{a}^{\;\;b}
\end{eqnarray}
for $(p,q)=(0,1),(1,0)$, 
\begin{eqnarray}
  \label{eq:second-Einstein-2,0-0,2}
  {}^{(p,q)}\!\bar{G}_{a}^{\;\;b}
  &=& 
  {}^{(1)}{\cal G}_{a}^{\;\;b}\left[{}^{(p,q)}{\cal H}\right]
  + {}^{(2)}{\cal G}_{a}^{\;\;b}
     \left[
       {}^{(p,q)}{\cal H}, {}^{(p,q)}{\cal H}
     \right]
  \nonumber\\
  && \quad
  + 2 {\pounds}_{\displaystyle {}^{(\frac{p}{2},\frac{q}{2})}X} \;\;
    {}^{(\frac{p}{2},\frac{q}{2})}\!\bar{G}_{a}^{\;\;b}
  + \left\{
    {\pounds}_{\displaystyle {}^{(p,q)}X}
    - {\pounds}_{\displaystyle {}^{(\frac{p}{2},\frac{q}{2})}X}^{2}
  \right\} G_{a}^{\;\;b}
\end{eqnarray}
for $(p,q)=(0,2),(2,0)$, and 
\begin{eqnarray}
  \label{eq:second-Einstein-1,1}
  {}^{(1,1)}\!\bar{G}_{a}^{\;\;b}
  &=& 
  {}^{(1)}{\cal G}_{a}^{\;\;b}\left[{}^{(1,1)}{\cal H}\right]
  + {}^{(2)}{\cal G}_{a}^{\;\;b}
     \left[
       {}^{(1,0)}{\cal H}, {}^{(0,1)}{\cal H}
     \right]
  \nonumber\\
  && \quad
  + {\pounds}_{\displaystyle {}^{(1,0)}_{}X} \;\; 
    {}^{(0,1)}\!\bar{G}_{a}^{\;\;b}
  + {\pounds}_{\displaystyle {}^{(0,1)}_{}X} \;\; 
    {}^{(1,0)}\!\bar{G}_{a}^{\;\;b}
  \nonumber\\
  && \quad
  + \left\{
    {\pounds}_{\displaystyle {}^{(1,1)}_{}X} 
    - \frac{1}{2} 
    {\pounds}_{\displaystyle {}^{(1,0)}_{}X} 
    {\pounds}_{\displaystyle {}^{(0,1)}_{}X} 
    - \frac{1}{2} 
    {\pounds}_{\displaystyle {}^{(0,1)}_{}X} 
    {\pounds}_{\displaystyle {}^{(1,0)}_{}X} 
  \right\} G_{a}^{\;\;b},
\end{eqnarray}
where
\begin{eqnarray}
  {}^{(1)}{\cal G}_{a}^{\;\;b}\left[A\right]
  &:=&
  - 2 \nabla_{[a}^{}H_{d]}^{\;\;\;bd}\left[A\right]
  - A^{cb} R_{ac}
  \nonumber\\
  && \quad
  + \frac{1}{2} \delta_{a}^{\;\;b}
  \left(
    2 \nabla_{[e}^{}H_{d]}^{\;\;\;ed}\left[A\right]
    + R_{ed} A^{ed}
  \right)
  \label{eq:cal-G-def-linear}
  , \\
  {}^{(2)}{\cal G}_{a}^{\;\;b}\left[A, B\right]
     &:=& 
    2 R_{ad} B_{c}^{\;\;(b}A^{d)c}
  + 2 H_{[a}^{\;\;\;de}\left[A\right] H_{d]\;\;e}^{\;\;\;b}\left[B\right]
  + 2 H_{[a}^{\;\;\;de}\left[B\right] H_{d]\;\;e}^{\;\;\;b}\left[A\right]
  \nonumber\\
  &&
  + 2 A_{e}^{\;\;d} \nabla_{[a}H_{d]}^{\;\;\;be}\left[B\right]
  + 2 B_{e}^{\;\;d} \nabla_{[a}H_{d]}^{\;\;\;be}\left[A\right]
  \nonumber\\
  &&
  + 2 A_{c}^{\;\;b} \nabla_{[a}H_{d]}^{\;\;\;cd}\left[B\right]
  + 2 B_{c}^{\;\;b} \nabla_{[a}H_{d]}^{\;\;\;cd}\left[A\right]
  \nonumber\\
  && \quad
  - \delta_{a}^{\;\;b}
  \left(
      R_{de} B_{f}^{\;\;(d} A^{e)f}
    + H_{[f}^{\;\;\;de}\left[A\right] H_{d]\;\;e}^{\;\;\;f}\left[B\right]
    + H_{[f}^{\;\;\;de}\left[B\right] H_{d]\;\;e}^{\;\;\;f}\left[A\right]
  \right.
  \nonumber\\
  && \quad\quad\quad\quad\quad
  \left.
    + 2 A_{e}^{\;\;d} \nabla_{[f}^{}H_{d]}^{\;\;\;[fe]}\left[B\right]
    + 2 B_{e}^{\;\;d} \nabla_{[f}^{}H_{d]}^{\;\;\;[fe]}\left[A\right]
  \right).
  \label{eq:cal-G-def-second}
\end{eqnarray}
We note that ${}^{(1)}{\cal G}_{a}^{\;\;b}\left[*\right]$ and 
${}^{(2)}{\cal G}_{a}^{\;\;b}\left[*,*\right]$ in
Eqs.~(\ref{eq:linear-Einstein})--(\ref{eq:second-Einstein-1,1})
are the gauge invariant parts of the perturbative Einstein tensors,
and each expression
(\ref{eq:linear-Einstein})--(\ref{eq:second-Einstein-1,1}) has
a form similar to
Eqs.~(\ref{eq:matter-gauge-inv-def-1.0})--(\ref{eq:matter-gauge-inv-def-1.1}),
respectively.


\subsection{Weyl curvature} 
\label{sec:weyl}

Here, we consider a perturbation of the Weyl curvature, which is 
useful to study some physical situations.
In $m$-dimensional spacetime, the Weyl curvature is defined by 
\begin{equation}
  \bar{C}_{abcd} := 
  \bar{R}_{abcd} 
  - \frac{2}{m-2} \left(\bar{g}_{a[c}\bar{R}_{d]b} 
    - \bar{g}_{b[c}\bar{R}_{d]a}\right)
  + \frac{2}{(m-1)(m-2)} \bar{R} \bar{g}_{a[c}\bar{g}_{d]b}.
  \label{eq:6.1}
\end{equation}
Using the perturbative formulae for each order perturbation of the
Riemann curvature
(\ref{eq:Riemann-dddd-linear})--(\ref{eq:Riemann-dddd-1,1}), of the
Ricci curvature
(\ref{eq:linear-order-gauge-inv-ricci})--(\ref{eq:epsilon-lambda-gauge-inv-ricci}),
of scalar curvature
(\ref{eq:linear-order-gauge-inv-scalar})--(\ref{eq:epsilon-lambda-gauge-inv-scalar}),
and of the metric perturbation
(\ref{eq:linear-metric-decomp})--(\ref{eq:H-ab-in-gauge-X-def-second-2}),     
we can explicitly confirm that the perturbative forms of the Weyl
curvature at each order are given by  
\begin{eqnarray}
  {}^{(p,q)}\!\bar{C}_{abcd} 
  &=& 
  {}^{(p,q)}\!{\cal C}_{abcd} 
  + {\pounds}_{{}^{(p,q)}\!X} C_{abcd}
  \quad\quad\quad
  \mbox{for} \quad
  (p,q) = (0,1), (1,0),
  \label{eq:linear-Weyl-dddd}
  \\
  {}^{(p,q)}\!\bar{C}_{abcd} 
  &=& 
  {}^{(p,q)}\!{\cal C}_{abcd} 
  + 2 {\pounds}_{{}^{(\frac{p}{2},\frac{q}{2})}\!X} 
  {}^{(\frac{p}{2},\frac{q}{2})}\!\bar{C}_{abcd}
  + \left(
    {\pounds}_{{}^{(p,q)}\!X}
    - {\pounds}_{{}^{(\frac{p}{2},\frac{q}{2})}\!X}^{2}
  \right)
  C_{abcd}
  \nonumber\\
  &&
  \quad\quad\quad\quad\quad
  \quad\quad\quad\quad\quad
  \quad\quad\quad
  \mbox{for} \quad
  (p,q) = (0,2), (2,0),
  \label{eq:second-Weyl-dddd}
  \\
  {}^{(1,1)}\!\bar{C}_{abcd} 
  &=& 
  {}^{(1,1)}\!{\cal C}_{abcd} 
  + {\pounds}_{{}^{(0,1)}\!X} {}^{(1,0)}\!\bar{C}_{abcd}
  + {\pounds}_{{}^{(1,0)}\!X} {}^{(0,1)}\!\bar{C}_{abcd}
  \nonumber\\
  &&
  \quad\quad
  + \left(
    {\pounds}_{{}^{(1,1)}\!X}
    - \frac{1}{2} {\pounds}_{{}^{(1,0)}\!X} {\pounds}_{{}^{(0,1)}\!X}
    - \frac{1}{2} {\pounds}_{{}^{(0,1)}\!X} {\pounds}_{{}^{(1,0)}\!X}
  \right)
  C_{abcd},
  \label{eq:1,1-second-Weyl-dddd}
\end{eqnarray}
where
\begin{eqnarray}
  {}^{(p,q)}\!{\cal C}_{abcd} 
  &=& 
  {}^{(p,q)}\!{\cal R}_{abcd}
  - \frac{2}{m-2} \left\{
      {}^{(p,q)}\!{\cal H}_{a[c} R_{d]b} 
    - {}^{(p,q)}\!{\cal H}_{b[c} R_{d]a}
  \right.
  \nonumber\\
  && \quad\quad\quad\quad\quad\quad\quad\quad\quad
  \left.
    + g_{a[c} {}^{(p,q)}\!{\cal R}_{d]b}
    - g_{b[c} {}^{(p,q)}\!{\cal R}_{d]a}
  \right\}
  \nonumber\\
  && \quad
  + \frac{2}{(m-1)(m-2)} 
  \left\{
    {}^{(p,q)}\!{\cal R} g_{a[c} g_{d]b}
    + R\;\; {}^{(p,q)}\!{\cal H}_{a[c} g_{d]b}
  \right.
  \nonumber\\
  && \quad\quad\quad\quad\quad\quad\quad\quad\quad\quad
  \left.
    + R\;\; g_{a[c} {}^{(p,q)}\!{\cal H}_{d]b}
  \right\}
  \label{eq:linear-Weyl-dddd-explicit}
\end{eqnarray}
for $(p,q) = (0,1), (1,0)$,
\begin{eqnarray}
  {}^{(p,q)}\!{\cal C}_{abcd} 
  &=& 
  {}^{(p,q)}\!{\cal R}_{abcd} 
  - \frac{2}{m-2} \left[
      {}^{(p,q)}\!{\cal H}_{a[c} R_{d]b} - {}^{(p,q)}\!{\cal H}_{b[c} R_{d]a}
  \right.
  \nonumber\\
  && \quad\quad\quad\quad\quad\quad\quad\quad\quad
  \left.
    + g_{a[c}\;{}^{(p,q)}\!{\cal R}_{d]b}
    - g_{b[c}\;{}^{(p,q)}\!{\cal R}_{d]a}
  \right.
  \nonumber\\
  && \quad\quad\quad\quad\quad\quad\quad\quad\quad
  \left.
    + 2 {}^{(\frac{p}{2},\frac{q}{2})}\!{\cal H}_{a[c}\;
        {}^{(\frac{p}{2},\frac{q}{2})}\!{\cal R}_{d]b}
    - 2 {}^{(\frac{p}{2},\frac{q}{2})}\!{\cal H}_{b[c}\;
        {}^{(\frac{p}{2},\frac{q}{2})}\!{\cal R}_{d]b}
  \right]
  \nonumber\\
  && \quad
  + \frac{2}{(m-1)(m-2)} 
  \left[
      {}^{(p,q)}\!
      {\cal R} g_{a[c} g_{d]b}
    + R\;{}^{(p,q)}\!{\cal H}_{a[c} g_{d]b}
    + R g_{a[c}\;{}^{(p,q)}\!{\cal H}_{d]b}
  \right.
  \nonumber\\
  && \quad\quad\quad\quad\quad\quad\quad\quad\quad\quad
  \left.
    + 2 {}^{(\frac{p}{2},\frac{q}{2})}\!{\cal R}
        g_{a[c}\;{}^{(\frac{p}{2},\frac{q}{2})}\!{\cal H}_{d]b}
    + 2 {}^{(\frac{p}{2},\frac{q}{2})}\!{\cal R}\;
        {}^{(\frac{p}{2},\frac{q}{2})}\!{\cal H}_{a[c} g_{d]b}
  \right.
  \nonumber\\
  && \quad\quad\quad\quad\quad\quad\quad\quad\quad\quad
  \left.
    + 2 R\;{}^{(\frac{p}{2},\frac{q}{2})}\!{\cal H}_{a[c}
         \;{}^{(\frac{p}{2},\frac{q}{2})}\!{\cal H}_{d]b}
  \right]
  \label{eq:second-Weyl-dddd-explicit}
\end{eqnarray}
for $(p,q) = (0,2), (2,0)$, and 
\begin{eqnarray}
  {}^{(1,1)}\!{\cal C}_{abcd} 
  &=& 
  {}^{(1,1)}\!{\cal R}_{abcd}
  - \frac{2}{m-2} \left[
      {}^{(1,1)}\!{\cal H}_{a[c} R_{d]b} 
    - {}^{(1,1)}\!{\cal H}_{b[c} R_{d]a}
  \right.
  \nonumber\\
  && \quad\quad\quad\quad\quad\quad\quad\quad\quad
  \left.
    + g_{a[c}\;{}^{(1,1)}\!{\cal R}_{d]b}
    - g_{b[c}\;{}^{(1,1)}\!{\cal R}_{d]a}
  \right.
  \nonumber\\
  && \quad\quad\quad\quad\quad\quad\quad\quad\quad
  \left.
    + {}^{(0,1)}\!{\cal H}_{a[c}\;{}^{(1,0)}\!{\cal R}_{d]b}
    + {}^{(1,0)}\!{\cal H}_{a[c}\;{}^{(0,1)}\!{\cal R}_{d]b}
  \right.
  \nonumber\\
  && \quad\quad\quad\quad\quad\quad\quad\quad\quad
  \left.
    - {}^{(0,1)}\!{\cal H}_{b[c}\;{}^{(1,0)}\!{\cal R}_{d]a}
    - {}^{(1,0)}\!{\cal H}_{b[c}\;{}^{(0,1)}\!{\cal R}_{d]a}
  \right]
  \nonumber\\
  &&
  + \frac{2}{(m-1)(m-2)} 
  \left[
      {}^{(1,1)}\!{\cal R} g_{a[c} g_{d]b}
    + R\;{}^{(1,1)}\!{\cal H}_{a[c} g_{d]b}
    + R g_{a[c}\;{}^{(1,1)}\!{\cal H}_{d]b}
  \right.
  \nonumber\\
  && \quad\quad\quad\quad\quad\quad\quad\quad\quad
  \left.
    + {}^{(0,1)}\!{\cal R}\;g_{a[c}\;{}^{(1,0)}\!{\cal H}_{d]b}
    + {}^{(0,1)}\!{\cal R}\;{}^{(1,0)}\!{\cal H}_{a[c}\;g_{d]b}
  \right.
  \nonumber\\
  && \quad\quad\quad\quad\quad\quad\quad\quad\quad
  \left.
    + {}^{(1,0)}\!{\cal R}\;{}^{(0,1)}\!{\cal H}_{a[c}\;g_{d]b}
    + {}^{(1,0)}\!{\cal R}\;g_{a[c}\;{}^{(0,1)}\!{\cal H}_{d]b}
  \right.
  \nonumber\\
  && \quad\quad\quad\quad\quad\quad\quad\quad\quad
  \left.
    + R\;{}^{(1,0)}\!{\cal H}_{a[c}{}^{(0,1)}\!{\cal H}_{d]b}
    + R\;{}^{(0,1)}\!{\cal H}_{a[c}{}^{(1,0)}\!{\cal H}_{d]b}
  \right].
  \label{eq:1,1-Weyl-dddd-explicit}
\end{eqnarray}


We also derive the formulae for the perturbative expansion of
$C_{abc}^{\;\;\;\;\;d}$ from 
\begin{eqnarray}
  \bar{C}_{abc}^{\;\;\;\;\;d} = \bar{g}^{ed}\bar{C}_{abce}.
\end{eqnarray}
Since the Weyl curvature is traceless, i.e.,
$\bar{C}_{adc}^{\;\;\;\;\;d}=0$, we can also verify the formulae
derived here by confirming this traceless property.


Actually, we explicitly confirm this traceless property with the
following formulae for the perturbative Weyl tensor at each order: 
\begin{eqnarray}
  {}^{(p,q)}\!\bar{C}_{abc}^{\;\;\;\;\;d} 
  &=& 
  {}^{(p,q)}\!{\cal C}_{abc}^{\;\;\;\;\;d}
  + {\pounds}_{{}^{(p,q)}\!X} C_{abc}^{\;\;\;\;\;d}
  \quad\quad\quad
  \mbox{for} \quad
  (p,q) = (0,1), (1,0),
  \label{eq:linear-Weyl-dddu}
  \\
  {}^{(p,q)}\!\bar{C}_{abc}^{\;\;\;\;\;d}
  &=& 
  {}^{(p,q)}\!{\cal C}_{abc}^{\;\;\;\;\;d}
  + 2 {\pounds}_{{}^{(\frac{p}{2},\frac{q}{2})}\!X} 
  {}^{(\frac{p}{2},\frac{q}{2})}\!\bar{C}_{abc}^{\;\;\;\;\;\;d}
  + \left(
    {\pounds}_{{}^{(p,q)}\!X}
    - {\pounds}_{{}^{(\frac{p}{2},\frac{q}{2})}\!X}^{2}
  \right)
  C_{abc}^{\;\;\;\;\;d}
  \nonumber\\
  &&
  \quad\quad\quad\quad\quad
  \quad\quad\quad\quad\quad
  \quad\quad\quad
  \mbox{for} \quad
  (p,q) = (0,2), (2,0),
  \label{eq:second-Weyl-dddu}
  \\
  {}^{(1,1)}\!\bar{C}_{abc}^{\;\;\;\;\;d}
  &=& 
  {}^{(1,1)}\!{\cal C}_{abc}^{\;\;\;\;\;d}
  + {\pounds}_{{}^{(0,1)}\!X} {}^{(1,0)}\!\bar{C}_{abc}^{\;\;\;\;\;\;d}
  + {\pounds}_{{}^{(1,0)}\!X} {}^{(0,1)}\!\bar{C}_{abc}^{\;\;\;\;\;\;d}
  \nonumber\\
  &&
  \quad\quad
  + \left(
    {\pounds}_{{}^{(1,1)}\!X}
    - \frac{1}{2} {\pounds}_{{}^{(1,0)}\!X} {\pounds}_{{}^{(0,1)}\!X}
    - \frac{1}{2} {\pounds}_{{}^{(0,1)}\!X} {\pounds}_{{}^{(1,0)}\!X}
  \right)
  C_{abc}^{\;\;\;\;\;d},
  \label{eq:1,1-second-Weyl-dddu}
\end{eqnarray}
where
\begin{eqnarray}
  {}^{(p,q)}\!{\cal C}_{abc}^{\;\;\;\;\;d}
  &:=& 
  -
  {}^{(p,q)}\!{\cal H}^{de} C_{abce} 
  +
  g^{de} {}^{(p,q)}\!\bar{C}_{abce}, 
  \label{eq:linear-Weyl-dddu-explicit}
  \\
  {}^{(p,q)}\!{\cal C}_{abc}^{\;\;\;\;\;d}
  &=& 
  - \left({}^{(p,q)}\!{\cal H}^{de} 
    - 2 {}^{(\frac{p}{2},\frac{q}{2})}\!{\cal H}_{f}^{\;\;e}\;
        {}^{(\frac{p}{2},\frac{q}{2})}\!{\cal H}^{df}\right)C_{abce}
  + g^{de}\;{}^{(p,q)}\!{\cal C}_{abce}
  \nonumber\\
  && \quad
  - 2 {}^{(\frac{p}{2},\frac{q}{2})}\!{\cal H}^{de}\;
      {}^{(\frac{p}{2},\frac{q}{2})}\!{\cal C}_{abce},
  \label{eq:second-Weyl-dddu-explicit}
  \\
  {}^{(1,1)}\!{\cal C}_{abc}^{\;\;\;\;\;d}
  &=& 
  - \left({}^{(1,1)}\!{\cal H}^{de} 
    - {}^{(1,0)}\!{\cal H}_{f}^{\;\;e}\;{}^{(0,1)}\!{\cal H}^{df}
    - {}^{(1,0)}\!{\cal H}_{f}^{\;\;d}\;{}^{(0,1)}\!{\cal H}^{ef}
  \right)C_{abce}
  \nonumber\\
  && \quad
  + g^{de}\;{}^{(1,1)}\!{\cal C}_{abce}
  - \;{}^{(0,1)}\!{\cal H}^{de}\;{}^{(1,0)}\!{\cal C}_{abce}
  - \;{}^{(1,0)}\!{\cal H}^{de}\;{}^{(0,1)}\!{\cal C}_{abce}.
  \label{eq:1,1-second-Weyl-dddu-explicit}
\end{eqnarray}
By using the fact that $C_{abc}^{\;\;\;\;\;b}=0$ for the Weyl
curvature on the background ${\cal M}_{0}$,
Eqs.~(\ref{eq:linear-Weyl-dddd-explicit}),
(\ref{eq:linear-order-perturbation-ricci}),
(\ref{eq:linear-order-perturbation-scalar}) and
(\ref{eq:Riemann-dddd-linear}), a straightforward calculation yields
\begin{eqnarray}
  {}^{(p,q)}\!{\cal C}_{abc}^{\;\;\;\;\;b} = 0
\end{eqnarray}
for $(p,q) = (0,1), (1,0)$.
Further, with straightforward calculations using
Eqs.~(\ref{eq:1,1-Weyl-dddd-explicit}), 
(\ref{eq:linear-Weyl-dddd-explicit}),
(\ref{eq:epsilon-lambda-gauge-inv-scalar}),
(\ref{eq:linear-order-perturbation-scalar}),
(\ref{eq:epsilon-lambda-gauge-inv-ricci}), 
(\ref{eq:linear-order-perturbation-ricci}),
(\ref{eq:Riemann-dddd-1,1}) and (\ref{eq:Riemann-dddd-linear}),
we can explicitly confirm the identity
\begin{eqnarray}
  {}^{(1,1)}\!{\cal C}_{abc}^{\;\;\;\;\;b} = 0,
\end{eqnarray}
and hence, we obtain
\begin{eqnarray}
  {}^{(p,q)}\!{\cal C}_{abc}^{\;\;\;\;\;b} = 0
\end{eqnarray}
for $(p,q) = (0,2), (2,0)$ through the replacements
(\ref{eq:replace-1,1-to-2,0}) and (\ref{eq:replace-1,1-to-0,2}) of the
perturbative variables. 
The property ${}^{(p,q)}\!{\cal C}_{abc}^{\;\;\;\;\;b} = 0$ is
trivial from the definition of the gauge invariant part of the
Weyl curvature. 
However, this trivial result gives us great confidence in the
formulae derived here.


\subsection{Divergence of an arbitrary tensor of second rank and the
  Bianchi identity}
\label{sec:bianchi-div-ener-momen-tensor}


Here, we consider the perturbation of the Bianchi identity and the
divergence of the energy momentum tensor, which are derived from the
divergence $\bar{\nabla}_{a}\bar{T}_{b}^{\;\;a}$ of an arbitrary
tensor field $\bar{T}_{b}^{\;\;a}$ of second rank.
The operator $\bar{\nabla}_{a}$ are the covariant derivative
associated with the metric $\bar{g}_{ab}$ on the physical
spacetime ${\cal M}$. 
As discussed above, $\bar{\nabla}_{a}$ is pulled back to
the background spacetime ${\cal M}_{0}$ as the derivative operator
${\cal X}^{*}\bar{\nabla}_{a}\left({\cal X}^{-1}\right)^{*}$ by
choosing a gauge ${\cal X}$. 
Further, the operation of $\bar{\nabla}_{a}$ as an operator on
${\cal M}_{0}$ is represented by the covariant derivative
$\nabla_{a}$, which is associated with the background metric
$g_{ab}$ on ${\cal M}_{0}$, and the tensor field $C^{c}_{\;\;ab}$
defined by Eq.~(\ref{eq:c-connection}).
Hence, we may concentrate on the Taylor expansion of the equation
\begin{equation}
  \bar{\nabla}_{a}\bar{T}_{b}^{\;\;a}
  = 
  \nabla_{a}\bar{T}_{b}^{\;\;a}
  + C^{a}_{\;\;ca}\bar{T}_{b}^{\;\;c}
  - C^{c}_{\;\;ba}\bar{T}_{c}^{\;\;a} 
  \label{eq:5.2}
\end{equation}
to derive the perturbative form of the divergence of an
arbitrary tensor field of second rank. 
The tensor field $\bar{T}_{a}^{\;\;b}$, which is pulled back
from the physical spacetime ${\cal M}_{\epsilon}$ to the
background spacetime ${\cal M}_{0}$, is expanded as
Eq.~(\ref{eq:expansion-form}). 
Following the definitions
(\ref{eq:matter-gauge-inv-def-1.0})--(\ref{eq:matter-gauge-inv-def-1.1})
of gauge invariant variables, the gauge invariant variables 
${}^{(p,q)}\!{\cal T}_{b}^{\;\;a}$ for each order perturbation
${}^{(p,q)}\!\bar{T}_{a}^{\;\;b}$ are defined by 
\begin{eqnarray}
  \label{eq:Tab-gauge-inv-def-1.0} 
  {}^{(p,q)}\!{\cal T}_{b}^{\;\;a} &:=&
  {}^{(p,q)}\!\bar{T}_{b}^{\;\;a} - {\pounds}_{{}^{(p,q)}X}T_{b}^{\;\;a},
  \quad\quad\quad\quad\quad\quad\quad
  (p,q) = (1,0), (0,1)
  , \\ 
  \label{eq:Tab-gauge-inv-def-2.0} 
  {}^{(p,q)}{\cal T}_{b}^{\;\;a} &:=&
  {}^{(p,q)}\!\bar{T}_{b}^{\;\;a} 
  - 2 {\pounds}_{{}^{(\frac{p}{2},\frac{q}{2})}X} 
  {}^{(\frac{p}{2},\frac{q}{2})}\!\bar{T}_{b}^{\;\;a} 
  \nonumber\\
  && \quad
  - \left\{
    {\pounds}_{{}^{(p,q)}X}
    -{\pounds}_{{}^{(\frac{p}{2},\frac{q}{2})}X}^{2}
  \right\} T_{b}^{\;\;a},
  \quad\quad\quad
  (p,q) = (2,0), (0,2)
  , \\
  \label{eq:Tab-gauge-inv-def-1.1} 
  {}^{(1,1)}{\cal T}_{b}^{\;\;a} &:=&
  {}^{(1,1)}\!\bar{T}_{b}^{\;\;a}
  - {\pounds}_{{}^{(1,0)}X} {}^{(0,1)}\!\bar{T}_{b}^{\;\;a}
  - {\pounds}_{{}^{(0,1)}X} {}^{(1,0)}\!\bar{T}_{b}^{\;\;a}
  \nonumber\\
  && \quad
  - \left\{{\pounds}_{{}^{(1,1)}X}
    - \frac{1}{2} {\pounds}_{{}^{(1,0)}X}
                  {\pounds}_{{}^{(0,1)}X}
    - \frac{1}{2} {\pounds}_{{}^{(0,1)}X}
                  {\pounds}_{{}^{(1,0)}X}
    \right\} T_{b}^{\;\;a}.
\end{eqnarray}
We also expand 
\begin{equation}
  \bar{\nabla}_{a}\bar{T}_{b}^{\;\;a}
  = 
  \sum_{k',k=0}^{\infty}
  \frac{\epsilon^{k}\lambda^{k'}}{k!k'!}
  \;
  {}^{(k,k')}\!\!
  \left(
    \bar{\nabla}_{a}\bar{T}_{b}^{\;\;a}
  \right).
\end{equation}


\subsubsection{Linear order}


A simple expansion of Eq.~(\ref{eq:5.2}) yields
\begin{eqnarray}
  {}^{(p,q)}\!\!
  \left(
    \bar{\nabla}_{a}\bar{T}_{b}^{\;\;a}
  \right)
  &=& 
  \nabla_{a}
  {}^{(p,q)}\!\bar{T}_{b}^{\;\;a}
  + 
  \left(
    H_{ca}^{\;\;\;\;a}\left[{}^{(p,q)}\!{\cal H}\right]
    + \nabla_{c}\nabla_{a}\;{}^{(p,q)}\!X^{a}
  \right)
  T_{b}^{\;\;c}
  \nonumber\\
  &&
  - 
  \left(
    H_{bac}\left[{}^{(p,q)}\!{\cal H}\right]
    + \nabla_{b}\nabla_{a}\;{}^{(p,q)}\!X_{c}
    + R_{acb}^{\;\;\;\;\;e}\;{}^{(p,q)}\!X_{e}
  \right)
  T^{ca}
\end{eqnarray}
for $(p,q)=(0,1),(1,0)$.
Then, using the gauge invariant variable defined by
Eq.~(\ref{eq:Tab-gauge-inv-def-1.0}), we obtain
\begin{eqnarray}
  {}^{(p,q)}\!\!
  \left(
    \bar{\nabla}_{a}\bar{T}_{b}^{\;\;a}
  \right)
  &=& 
  \nabla_{a}
  {}^{(p,q)}\!{\cal T}_{b}^{\;\;a}
  + H_{ca}^{\;\;\;\;a}\left[{}^{(p,q)}\!{\cal H}\right] T_{b}^{\;\;c}
  - H_{ba}^{\;\;\;\;c}\left[{}^{(p,q)}\!{\cal H}\right] T_{c}^{\;\;a}
  \nonumber\\
  &&
  + {\pounds}_{{}^{(p,q)}\!X}\nabla_{a}T_{b}^{\;\;a}
  \label{eq:linear-order-divergence-of-Tab}
\end{eqnarray}
for linear order [$(p,q)=(1,0), (0,1)$], where we have used the
formula (\ref{eq:5.13}).


Now, we check the linear-order perturbation of the Bianchi
identity by using Eq.~(\ref{eq:linear-order-divergence-of-Tab}). 
To do this, we regard $T_{a}^{\;\;b}$ as the Einstein tensor
$G_{a}^{\;\;b}$.
Further, the gauge invariant part of the tensor
$T_{a}^{\;\;b}$ is regarded as the gauge invariant part of the linear
order Einstein tensor:
\begin{eqnarray}
  T_{a}^{\;\;b} &\rightarrow& G_{a}^{\;\;b}, \\
  {}^{(p,q)}\!{\cal T}_{a}^{\;\;b} 
  &\rightarrow& 
  {}^{(1)}{\cal G}_{a}^{\;\;b}\left[{}^{(p,q)}\!{\cal H}\right].
\end{eqnarray}
Through these replacements, we obtain 
\begin{eqnarray}
  {}^{(p,q)}\!\!
  \left(
    \bar{\nabla}_{a}\bar{G}_{b}^{\;\;a}
  \right)
  &=& 
  \nabla_{a}
  {}^{(1)}{\cal G}_{a}^{\;\;b}\left[{}^{(p,q)}\!{\cal H}\right]
  + H_{ca}^{\;\;\;\;a}\left[{}^{(p,q)}\!{\cal H}\right] G_{b}^{\;\;c}
  - H_{ba}^{\;\;\;\;c}\left[{}^{(p,q)}\!{\cal H}\right] G_{c}^{\;\;a}
  \nonumber\\
  &&
  + {\pounds}_{{}^{(p,q)}\!X}\nabla_{a}G_{b}^{\;\;a}.
  \label{eq:linear-order-divergence-of-Gab}
\end{eqnarray}
On the other hand, a direct calculation using
Eq.~(\ref{eq:cal-G-def-linear}) yields 
\begin{eqnarray}
  \nabla_{a}
  {}^{(1)}{\cal G}_{b}^{\;\;a}\left[{}^{(p,q)}\!{\cal H}\right]
  &=& 
  - H_{ca}^{\;\;\;\;a}\left[{}^{(p,q)}\!{\cal H}\right] G_{b}^{\;\;c}
  + H_{ba}^{\;\;\;\;c}\left[{}^{(p,q)}\!{\cal H}\right] G_{c}^{\;\;a}
  \label{eq:linear-order-divergence-of-calGab}
\end{eqnarray}
as an identity.
Therefore, the linear-order expansion of the divergence of the
Einstein tensor is given by 
\begin{eqnarray}
  {}^{(p,q)}\!\!
  \left(
    \bar{\nabla}_{a}\bar{G}_{b}^{\;\;a}
  \right)
  &=& 
  {\pounds}_{{}^{(p,q)}\!X}\nabla_{a}G_{b}^{\;\;a}, 
  \quad
  (p,q)=(1,0), (0,1).
\end{eqnarray}
Because $\nabla_{a}G_{b}^{\;\;a}=0$ for an arbitrary spacetime, we can 
easily see that 
${}^{(p,q)}\!\!\left(\bar{\nabla}_{a}\bar{G}_{b}^{\;\;a}\right)=0$,
identically, at linear order.


\subsubsection{Second order}


Next, we consider the $O(\epsilon\lambda)$ perturbation of
$\nabla_{a}{T}_{b}^{\;\;a}$. 
A straightforward calculations yields 
\begin{eqnarray}
  {}^{(1,1)}\!\!
  \left(
    \bar{\nabla}_{a}\bar{T}_{b}^{\;\;a}
  \right)
  &=& 
  \nabla_{a}
  {}^{(1,1)}\!{\cal T}_{b}^{\;\;a}
  \nonumber\\
  && 
  - 
  \left(
      H_{cad}\left[{}^{(0,1)}\!{\cal H}\right] {}^{(1,0)}\!{\cal H}^{da}
    + H_{cad}\left[{}^{(1,0)}\!{\cal H}\right] {}^{(0,1)}\!{\cal H}^{da}
    - H_{ca}^{\;\;\;\;a}\left[{}^{(1,1)}\!{\cal H}\right]
  \right)
  T_{b}^{\;\;c}
  \nonumber\\
  && 
  + 
  \left(
      H_{bad}\left[{}^{(0,1)}\!{\cal H}\right] {}^{(1,0)}\!{\cal H}^{dc}
    + H_{bad}\left[{}^{(1,0)}\!{\cal H}\right] {}^{(0,1)}\!{\cal H}^{dc}
    - H_{ba}^{\;\;\;\;c}\left[{}^{(1,1)}\!{\cal H}\right]
  \right)
  T_{c}^{\;\;a}
  \nonumber\\
  && 
  - H_{ba}^{\;\;\;\;c}\left[{}^{(1,0)}\!{\cal H}\right]
    {}^{(0,1)}\!{\cal T}_{c}^{\;\;a}
  + H_{ca}^{\;\;\;\;a}\left[{}^{(1,0)}\!{\cal H}\right]
    {}^{(0,1)}\!{\cal T}_{b}^{\;\;c}
  \nonumber\\
  && 
  - H_{ba}^{\;\;\;\;c}\left[{}^{(0,1)}\!{\cal H}\right]
    {}^{(1,0)}\!{\cal T}_{c}^{\;\;a}
  + H_{ca}^{\;\;\;\;a}\left[{}^{(0,1)}\!{\cal H}\right]
    {}^{(1,0)}\!{\cal T}_{b}^{\;\;c}
  \nonumber\\
  && 
  + {\pounds}_{{}^{(1,0)}\!X} \;
  {}^{(0,1)}\!\!
  \left(
    \bar{\nabla}_{a}\bar{T}_{b}^{\;\;a}
  \right)
  + {\pounds}_{{}^{(0,1)}\!X} \;
  {}^{(1,0)}\!\!
  \left(
    \bar{\nabla}_{a}\bar{T}_{b}^{\;\;a}
  \right)
  \nonumber\\
  && 
  + \left\{
    {\pounds}_{{}^{(1,1)}\!X}
    - \frac{1}{2} {\pounds}_{{}^{(0,1)}\!X} {\pounds}_{{}^{(1,0)}\!X}
    - \frac{1}{2} {\pounds}_{{}^{(1,0)}\!X} {\pounds}_{{}^{(0,1)}\!X}
  \right\} \;
  \left(\nabla_{a}T_{b}^{\;\;a}\right).
  \label{eq:second-order-div-tmunu-1-1}
\end{eqnarray}
Applying the replacements (\ref{eq:replace-1,1-to-2,0}) and
(\ref{eq:replace-1,1-to-0,2}) of the perturbative variables to
Eq.~(\ref{eq:second-order-div-tmunu-1-1}), 
the $O(\epsilon^{2})$ and $O(\lambda^{2})$ perturbations of the
divergence of a tensor $T_{a}^{\;\;b}$ are obtained as
\begin{eqnarray}
  {}^{(p,q)}\!\!
  \left(
    \bar{\nabla}_{a}\bar{T}_{b}^{\;\;a}
  \right)
  &=& 
  \nabla_{a}
  {}^{(p,q)}\!{\cal T}_{b}^{\;\;a}
  \nonumber\\
  && 
  - 
  \left(
    2 H_{cad} \left[{}^{(\frac{p}{2},\frac{q}{2})}\!{\cal H}\right] \;
    {}^{(\frac{p}{2},\frac{q}{2})}\!{\cal H}^{da}
    - H_{ca}^{\;\;\;\;a}\left[{}^{(p,q)}\!{\cal H}\right]
  \right)
  T_{b}^{\;\;c}
  \nonumber\\
  && 
  + 
  \left(
    2 H_{bad} \left[{}^{(\frac{p}{2},\frac{q}{2})}\!{\cal H}\right] \;
    {}^{(\frac{p}{2},\frac{q}{2})}\!{\cal H}^{dc}
    - H_{ba}^{\;\;\;\;c}\left[{}^{(p,q)}\!{\cal H}\right]
  \right)
  T_{c}^{\;\;a}
  \nonumber\\
  && 
  - 2 H_{ba}^{\;\;\;\;c}\left[{}^{(\frac{p}{2},\frac{q}{2})}\!{\cal H}\right]\;
      {}^{(\frac{p}{2},\frac{q}{2})}\!{\cal T}_{c}^{\;\;a}
  + 2 H_{ca}^{\;\;\;\;a}\left[{}^{(\frac{p}{2},\frac{q}{2})}\!{\cal H}\right]\;
      {}^{(\frac{p}{2},\frac{q}{2})}\!{\cal T}_{b}^{\;\;c}
  \nonumber\\
  && 
  + 2 {\pounds}_{{}^{(\frac{p}{2},\frac{q}{2})}\!X} \;
      {}^{(\frac{p}{2},\frac{q}{2})}\!\!
      \left(
        \bar{\nabla}_{a}\bar{T}_{b}^{\;\;a}
      \right)
  + \left\{
    {\pounds}_{{}^{(p,q)}\!X}
    - {\pounds}_{{}^{(\frac{p}{2},\frac{q}{2})}\!X}^{2}
  \right\} \;
  \left(\nabla_{a}T_{b}^{\;\;a}\right),
  \label{eq:3.97}
\end{eqnarray}
where $(p,q)=(2,0), (0,2)$.


As in the linear-order case, we can also check the Bianchi
identities of the second order perturbations using Eqs.~(\ref{eq:3.97})
and (\ref{eq:second-order-div-tmunu-1-1}).
To do this, we regard $T_{a}^{\;\;b}$ as the Einstein tensor
$G_{a}^{\;\;b}$ and apply the following replacement:
\begin{eqnarray}
  T_{a}^{\;\;b} &\rightarrow& G_{a}^{\;\;b}, \\
  {}^{(p,q)}\!{\cal T}_{a}^{\;\;b}
  &\rightarrow&
  {}^{(1)}\!{\cal G}_{a}^{\;\;b}\left[{}^{(p,q)}\!{\cal H}\right]
  \quad\quad \mbox{for} \quad
  (p,q) = (1,0), (0,1),
  \\ 
  {}^{(p,q)}\!{\cal T}_{a}^{\;\;b}
  &\rightarrow&
  {}^{(1)}\!{\cal G}_{a}^{\;\;b}\left[{}^{(p,q)}\!{\cal H}\right]
  + 
  {}^{(2)}\!{\cal G}_{a}^{\;\;b}\left[
    {}^{(\frac{p}{2},\frac{q}{2})}\!{\cal H},
    {}^{(\frac{p}{2},\frac{q}{2})}\!{\cal H}
  \right]
  \nonumber\\
  && 
  \quad\quad\quad\quad\quad\quad\quad\quad\mbox{for}\quad
  (p,q) = (2,0), (0,2),
  \\ 
  {}^{(1,1)}\!{\cal T}_{a}^{\;\;b}
  &\rightarrow&
  {}^{(1)}\!{\cal G}_{a}^{\;\;b}\left[{}^{(1,1)}\!{\cal H}\right]
  + 
  {}^{(2)}\!{\cal G}_{a}^{\;\;b}\left[
    {}^{(1,0)}\!{\cal H},
    {}^{(0,1)}\!{\cal H}
  \right].
\end{eqnarray}
First, we consider the perturbative Bianchi identity of
$O(\epsilon\lambda)$:
\begin{eqnarray}
  {}^{(1,1)}\!\!
  \left(
    \bar{\nabla}_{a}\bar{G}_{b}^{\;\;a}
  \right)
  &=& 
  \nabla_{a}
  \left(
    {}^{(1)}\!{\cal G}_{b}^{\;\;a}
    \left[
      {}^{(1,1)}\!{\cal H}
    \right]
    + 
    {}^{(2)}\!{\cal G}_{b}^{\;\;a}
    \left[
      {}^{(1,0)}\!{\cal H},
      {}^{(0,1)}\!{\cal H}
    \right]
  \right)
  \nonumber\\
  && 
  - 
  \left(
      H_{cad}\left[{}^{(0,1)}\!{\cal H}\right] \;
      {}^{(1,0)}\!{\cal H}^{da}
    + H_{cad}\left[{}^{(1,0)}\!{\cal H}\right] \;
      {}^{(0,1)}\!{\cal H}^{da}
    - H_{ca}^{\;\;\;\;a}\left[{}^{(1,1)}\!{\cal H}\right]
  \right)
  G_{b}^{\;\;c}
  \nonumber\\
  && 
  + 
  \left(
      H_{bad}\left[{}^{(0,1)}\!{\cal H}\right] \; {}^{(1,0)}\!{\cal H}^{dc}
    + H_{bad}\left[{}^{(1,0)}\!{\cal H}\right] \; {}^{(0,1)}\!{\cal H}^{dc}
    - H_{ba}^{\;\;\;\;c}\left[{}^{(1,1)}\!{\cal H}\right]
  \right)
  G_{c}^{\;\;a}
  \nonumber\\
  && 
  - H_{ba}^{\;\;\;\;c}\left[{}^{(1,0)}\!{\cal H}\right]\;
    {}^{(0,1)}\!{\cal G}_{c}^{\;\;a}
  + H_{ca}^{\;\;\;\;a}\left[{}^{(1,0)}\!{\cal H}\right]\;
    {}^{(0,1)}\!{\cal G}_{b}^{\;\;c}
  \nonumber\\
  && 
  - H_{ba}^{\;\;\;\;c}\left[{}^{(0,1)}\!{\cal H}\right]\;
    {}^{(1,0)}\!{\cal G}_{c}^{\;\;a}
  + H_{ca}^{\;\;\;\;a}\left[{}^{(0,1)}\!{\cal H}\right]\;
    {}^{(1,0)}\!{\cal G}_{b}^{\;\;c}
  \nonumber\\
  && 
  + {\pounds}_{{}^{(1,0)}\!X} \;
  {}^{(0,1)}\!\!
  \left(
    \bar{\nabla}_{a}\bar{G}_{b}^{\;\;a}
  \right)
  + {\pounds}_{{}^{(0,1)}\!X} \;
  {}^{(1,0)}\!\!
  \left(
    \bar{\nabla}_{a}\bar{G}_{b}^{\;\;a}
  \right)
  \nonumber\\
  && 
  + \left\{
    {\pounds}_{{}^{(1,1)}\!X}
    - \frac{1}{2} {\pounds}_{{}^{(0,1)}\!X} {\pounds}_{{}^{(1,0)}\!X}
    - \frac{1}{2} {\pounds}_{{}^{(1,0)}\!X} {\pounds}_{{}^{(0,1)}\!X}
  \right\} \;
  \left(\nabla_{a}G_{b}^{\;\;a}\right).
  \label{eq:second-order-div-bianchi-11-org}
\end{eqnarray}
Here, we note that the identity
(\ref{eq:linear-order-divergence-of-calGab}) implies
\begin{eqnarray}
  \nabla_{a} {}^{(1)}{\cal G}_{b}^{\;\;a}\left[{}^{(1,1)}\!{\cal H}\right]
  = 
  - H_{ca}^{\;\;\;\;a}
  \left[
    {}^{(1,1)}\!{\cal H}
  \right]
  G_{b}^{\;\;c}
  + H_{ba}^{\;\;\;\;c}
  \left[
    {}^{(1,1)}\!{\cal H}
  \right]
  G_{c}^{\;\;a}.
  \label{eq:linear-div-of-calGab-1,1}
\end{eqnarray}
Further, straightforward calculations lead to the following
identity: 
\begin{eqnarray}
  && 
  \nabla_{a}{}^{(2)}{\cal G}_{b}^{\;\;a}
  \left[
    {}^{(1,0)}\!{\cal H},
    {}^{(0,1)}\!{\cal H}
  \right] 
  \nonumber\\
  &=& 
  -   H_{ca}^{\;\;\;\;a}
  \left[
    {}^{(1,0)}\!{\cal H}
  \right]
  \; {}^{(1)}\!{\cal G}_{b}^{\;\;c}
  \left[
    {}^{(0,1)}\!{\cal H}
  \right]
  -   H_{ca}^{\;\;\;\;a}
  \left[
    {}^{(0,1)}\!{\cal H}
  \right]
  \; {}^{(1)}\!{\cal G}_{b}^{\;\;c}
  \left[
    {}^{(1,0)}\!{\cal H}
  \right]
  \nonumber\\
  &&
  +   H_{ba}^{\;\;\;\;e}
  \left[
    {}^{(1,0)}\!{\cal H}
  \right]
  \; {}^{(1)}\!{\cal G}_{e}^{\;\;a}
  \left[
    {}^{(0,1)}\!{\cal H}
  \right]
  +   H_{ba}^{\;\;\;\;e}
  \left[
    {}^{(0,1)}\!{\cal H}
  \right]
  \; {}^{(1)}\!{\cal G}_{e}^{\;\;a}
  \left[
    {}^{(1,0)}\!{\cal H}
  \right]
  \nonumber\\
  &&
  - \left(
    H_{bad}
    \left[
      {}^{(0,1)}\!{\cal H}
    \right]
    \; {}^{(1,0)}\!{\cal H}^{dc}
    +
    H_{bad}
    \left[
      {}^{(1,0)}\!{\cal H}
    \right]
    \; {}^{(0,1)}\!{\cal H}^{dc}
  \right)
  G_{c}^{\;\;a}
  \nonumber\\
  &&
  + \left(
    H_{cad}
    \left[
      {}^{(0,1)}\!{\cal H}
    \right]
    \; {}^{(1,0)}\!{\cal H}^{ad}
    +
    H_{cad}
    \left[
      {}^{(1,0)}\!{\cal H}
    \right]
    \; {}^{(0,1)}\!{\cal H}^{ad}
  \right)
  G_{b}^{\;\;c}.
  \label{eq:second-div-of-calGab-1,1}
\end{eqnarray}
Using the identities
(\ref{eq:linear-div-of-calGab-1,1})
and
(\ref{eq:second-div-of-calGab-1,1})
, we easily find
\begin{eqnarray}
  {}^{(1,1)}\!\!
  \left(
    \bar{\nabla}_{a}\bar{G}_{b}^{\;\;a}
  \right)
  &&= 
  {\pounds}_{{}^{(1,0)}\!X} \;
  {}^{(0,1)}\!\!
  \left(
    \bar{\nabla}_{a}\bar{G}_{b}^{\;\;a}
  \right)
  + {\pounds}_{{}^{(0,1)}\!X} \;
  {}^{(1,0)}\!\!
  \left(
    \bar{\nabla}_{a}\bar{G}_{b}^{\;\;a}
  \right)
  \nonumber\\
  && 
  + \left\{
    {\pounds}_{{}^{(1,1)}\!X}
    - \frac{1}{2} {\pounds}_{{}^{(0,1)}\!X} {\pounds}_{{}^{(1,0)}\!X}
    - \frac{1}{2} {\pounds}_{{}^{(1,0)}\!X} {\pounds}_{{}^{(0,1)}\!X}
  \right\}
  \left(\nabla_{a}G_{b}^{\;\;a}\right).
  \label{eq:second-order-div-explicit-1-1}
\end{eqnarray}
Since the Bianchi identities $\nabla_{a}G_{b}^{\;\;a}=0$ on the
background spacetime ${\cal M}_{0}$ and those of the linear-order
perturbations ${}^{(p,q)}\!(\nabla_{a}G_{b}^{\;\;a})=0$
[for $(p,q)=(1,0),(0,1)$] have already been confirmed, we have also
confirmed the identity 
\begin{eqnarray}
  {}^{(1,1)}\!\!
  \left(
    \bar{\nabla}_{a}\bar{G}_{b}^{\;\;a}
  \right)
  &=& 0.
\end{eqnarray}


Applying the replacements (\ref{eq:replace-1,1-to-2,0}) and
(\ref{eq:replace-1,1-to-0,2}) of the perturbative variables to
Eq.~(\ref{eq:second-order-div-bianchi-11-org}), we obtain the
$O(\epsilon^{2})$ and $O(\lambda^{2})$ Bianchi identities:
\begin{eqnarray}
  {}^{(p,q)}\!\!
  \left(
    \bar{\nabla}_{a}\bar{G}_{b}^{\;\;a}
  \right)
  &=& 
  \nabla_{a}
  \left(
    {}^{(1)}\!{\cal G}_{b}^{\;\;a}
    \left[
      {}^{(p,q)}\!{\cal H}
    \right]
    + 
    {}^{(2)}\!{\cal G}_{b}^{\;\;a}
    \left[
      {}^{(\frac{p}{2},\frac{q}{2})}\!{\cal H},
      {}^{(\frac{p}{2},\frac{q}{2})}\!{\cal H}
    \right]
  \right)
  \nonumber\\
  && 
  - 
  \left(
    2 H_{cad}\left[{}^{(\frac{p}{2},\frac{q}{2})}\!{\cal H}\right] \;
    {}^{(\frac{p}{2},\frac{q}{2})}\!{\cal H}^{da}
    - H_{ca}^{\;\;\;\;a}\left[{}^{(p,q)}\!{\cal H}\right]
  \right)
  G_{b}^{\;\;c}
  \nonumber\\
  && 
  + 
  \left(
    2 H_{bad}\left[{}^{(\frac{p}{2},\frac{q}{2})}\!{\cal H}\right] \;
    {}^{(\frac{p}{2},\frac{q}{2})}\!{\cal H}^{dc}
    - H_{ba}^{\;\;\;\;c}\left[{}^{(p,q)}\!{\cal H}\right]
  \right)
  G_{c}^{\;\;a}
  \nonumber\\
  && 
  - 2 H_{ba}^{\;\;\;\;c}\left[{}^{(\frac{p}{2},\frac{q}{2})}\!{\cal H}\right]\;
  {}^{(1)}\!{\cal G}_{c}^{\;\;a}
  \left[
    {}^{(\frac{p}{2},\frac{q}{2})}\!{\cal H}
  \right]
  + 2 H_{ca}^{\;\;\;\;a}\left[{}^{(\frac{p}{2},\frac{q}{2})}\!{\cal H}\right]\;
  {}^{(1)}\!{\cal G}_{b}^{\;\;c}
  \left[
    {}^{(\frac{p}{2},\frac{q}{2})}\!{\cal H}
  \right]
  \nonumber\\
  && 
  + 2 {\pounds}_{{}^{(\frac{p}{2},\frac{q}{2})}\!X} \;
  {}^{(\frac{p}{2},\frac{q}{2})}\!\!
  \left(
    \bar{\nabla}_{a}\bar{G}_{b}^{\;\;a}
  \right)
  + \left\{
    {\pounds}_{{}^{(p,q)}\!X}
    - {\pounds}_{{}^{(\frac{p}{2},\frac{q}{2})}\!X}^{2}
  \right\} \;
  \left(\nabla_{a}G_{b}^{\;\;a}\right)
  \nonumber\\
  &=& 0,
  \label{eq:second-order-div-20-02}
\end{eqnarray}
where $(p,q)=(0,2), (2,0)$.

\subsection{Einstein equations} 
\label{sec:Einstein-eq}

Finally, we consider perturbations of the Einstein equation
at each order.
First, we expand the energy momentum tensor as
Eq.~(\ref{eq:expansion-form}) and impose the perturbed
Einstein equation of each order,
\begin{equation}
  {}^{(p,q)}G_{a}^{\;\;b} = 8\pi G \;\; {}^{(p,q)}T_{a}^{\;\;b}.
\end{equation}
Then, we define the gauge invariant variable 
${}^{(p,q)}{\cal T}_{a}^{\;\;b}$ for the perturbative energy
momentum tensor at each order by 
Eqs.~(\ref{eq:Tab-gauge-inv-def-1.0})--(\ref{eq:Tab-gauge-inv-def-1.1}).
Then, the perturbative Einstein equation at each order is given by 
\begin{eqnarray}
  \label{eq:each-order-perturbations}
  {}^{(1)}{\cal G}_{a}^{\;\;b}\left[{}^{(p,q)}{\cal H}\right]
  &=&
  8\pi G \;\; {}^{(p,q)}{\cal T}_{a}^{\;\;b}
\end{eqnarray}
for linear order [$(p,q)=(0,1),(1,0)$] and
\begin{eqnarray}
  {}^{(1)}{\cal G}_{a}^{\;\;b}\left[{}^{(1,1)}{\cal H}\right]
  + {}^{(2)}{\cal G}_{a}^{\;\;b}
     \left[
       {}^{(1,0)}{\cal H},
       {}^{(0,1)}{\cal H}
     \right]
  &=&
  8\pi G \;\; {}^{(1,1)}{\cal T}_{a}^{\;\;b} 
  , \\
  {}^{(1)}{\cal G}_{a}^{\;\;b}\left[{}^{(p,q)}{\cal H}\right]
  + {}^{(2)}{\cal G}_{a}^{\;\;b}
     \left[
       {}^{(\frac{p}{2},\frac{q}{2})}{\cal H},
       {}^{(\frac{p}{2},\frac{q}{2})}{\cal H}
     \right]
  &=&
  8\pi G \;\; {}^{(p,q)}{\cal T}_{a}^{\;\;b}
\end{eqnarray}
for second order [$(p,q)=(0,2),(2,0)$].
These explicitly show that, order by order, the Einstein
equations are necessarily obtained in terms of gauge invariant
variables only.


\section{Summary and Discussions}
\label{sec:summary}


In summary, we have derived some formulae for second-order gauge
invariant perturbations, namely those of the Riemann, Ricci,
scalar, Einstein, and Weyl curvature tensors. 
We also derived the formulae for the divergence of an arbitrary
tensor field of second rank for perturbations at each order. 
These perturbative curvatures have the same forms as those in
the definitions
(\ref{eq:matter-gauge-inv-def-1.0})--(\ref{eq:matter-gauge-inv-def-1.1})
of the gauge invariant variables for arbitrary perturbative
fields which are proposed in a previous paper\cite{kouchan-gauge-inv}.
These are useful for investigating physical problems using
perturbation theory.


Through the derivation of these formulae, we have confirmed some
facts.
First, if linear-order gauge invariant perturbation theory
is well established, its extension to higher orders and
multi-parameter perturbations is straightforward.
Second, the perturbative Weyl curvature at each order preserve 
the property of the tracelessness of the Weyl curvature. 
Third, the perturbative Einstein curvature of each order satisfy
the perturbative Bianchi identities at each order. 
Fourth, the perturbative Einstein equations at each order 
necessarily take in gauge invariant forms.
These properties we have confirmed are trivial from their
derivations. 
In particular, the fourth result is trivial, because any equation
can be written in a form in which the right-hand side is equal to
``zero'' in any gauge. 
This ``zero'' is gauge invariant.
Hence, the left-hand side should be gauge invariant.
However, we emphasize that these trivial results imply that
the formulae derived here and the framework of higher-order
perturbation theory applied here are mathematically consistent
at this level.


Further, we note that in our framework, we specify nothing about
the background spacetime nor about the physical meaning of the
parameters for the perturbations. 
Our framework is based only on general covariance.
For this reason, this framework is applicable to any theory in
which general covariance is imposed, and thus it has very many
applications. 
Actually, we are planning to apply it to some physical problems.
The following are candidates of the physical situations
to which the second order perturbation theory should be applied:
the radiation reaction problem in gravitational wave
emission\cite{radiation-reaction};
stationary axisymmetric ideal MHD flow around a black hole or a
star\cite{MHD-ref};
the correspondence between observables in experiment and gauge
invariant variables;
dynamics of gravitating membranes (for example, topological
defects\cite{kouchan-papers}, brane world\cite{brane-papers}, and
so on);
perturbations of a compact star with rotation and pulsation\cite{Kojima};
Post-Minkowski expansion alternative to post-Newtonian
expansion\cite{Post-Minkowski-Detweiler}; 
higher-order cosmological perturbations and primordial
non-Gaussianity\cite{cosmological-second-non-gauss}.


In particular, the gauge invariant form of the second-order
perturbation of the divergence of the energy momentum tensor
should be useful in considering the gauge problem in the
radiation reaction of gravitational wave emission. 
In astrophysical contexts, it is natural to consider the
situation in which a solar mass object falls into a supermassive
black hole of mass $\sim 10^{6}M_{\odot}$. 
This is one of the target phenomena of the observation of
gravitational wave by LISA (laser interferometer space antenna for
gravitational wave measurements)\cite{LISA-home-page}.
In such a situation, the perturbation parameter is the ratio of
the mass of the compact object to that of the central
supermassive black hole. 
We may regard the second order perturbations as describing the
radiation reaction effect of gravitational wave emission. 
By applying the gauge invariant formulation discussed here, 
we can exclude gauge freedom completely and thus there is no gauge
ambiguities in the results.
It would be quite interesting to apply the formulation discussed here
to this radiation reaction problem. 
We leave this application as a future work.


In addition to the radiation reaction problem of gravitational wave
emission, there are many applications, some of which are listed
above. 
Indeed, because there are few assumptions in our treatment, it
is natural to expect that there are many applications that are
not listed above.  
We also expect that the formulae derived here will be found to
be very powerful tools in many applications.


\section*{Acknowledgements}
The author acknowledges Prof. L.~Kofman and Prof. R.M.~Wald and
their colleagues for the hospitality during my visit to CITA and
the Relativity Group at the University of Chicago.
The author also thanks to Prof. M.~Sasaki for the valuable
comments on the title of this paper.
Finally, the author deeply thanks all members of Division of
Theoretical Astronomy at NAOJ and other colleagues for their
continuous encouragement.


\appendix
\section{Useful formulae}
\label{sec:use-ful}

In the calculations in the main text, the knowledge of the
commutation relations of the covariant derivative and Lie
derivative are useful.
Here, we summarize the commutation relations which are used in
the derivation in the main text:
\begin{eqnarray}
  \nabla_{a} {\pounds}_{X} t_{b}
  &=&
  {\pounds}_{X}\nabla_{a}t_{b} 
  + X^{c}R_{acb}^{\;\;\;\;\;\;d}t_{d}
  + t_{c}\nabla_{a}\nabla_{b}X^{c}
  , \\
  \nabla_{a} {\pounds}_{X} t_{bc}
  &=&
    {\pounds}_{X}\nabla_{a}t_{bc} 
  + X^{d}R_{adb}^{\;\;\;\;\;\;e}t_{ec} 
  + X^{d}R_{adc}^{\;\;\;\;\;\;e}t_{be} 
  \nonumber\\
  && \quad
  + t_{dc}\nabla_{a}\nabla_{b}X^{d} 
  + t_{bd}\nabla_{a}\nabla_{c}X^{d}
  , \\
  \nabla_{a} {\pounds}_{X} t_{b}^{\;\;c}
  &=&
    {\pounds}_{X}\nabla_{a}t_{b}^{\;\;c}
  + X^{d}R_{adb}^{\;\;\;\;\;\;e} t_{e}^{\;\;c}
  - X^{d}R_{ade}^{\;\;\;\;\;\;c} t_{b}^{\;\;e}
  \nonumber\\
  && \quad
  + t_{d}^{\;\;c} \nabla_{a}\nabla_{b}X^{d}
  - t_{b}^{\;\;d} \nabla_{a}\nabla_{d}X^{c},
  \\
  \label{eq:5.13}
  \nabla_{a} {\pounds}_{X} t_{bcd}
  &=&
    {\pounds}_{X}\nabla_{a}t_{bcd} 
  + X^{e}R_{aeb}^{\;\;\;\;\;\;f}t_{fcd} 
  + X^{e}R_{aec}^{\;\;\;\;\;\;f}t_{bfd} 
  + X^{e}R_{aed}^{\;\;\;\;\;\;f}t_{bcf} 
  \nonumber\\
  && \quad
  + t_{ecd}\nabla_{a}\nabla_{b}X^{e} 
  + t_{bed}\nabla_{a}\nabla_{c}X^{e} 
  + t_{bce}\nabla_{a}\nabla_{d}X^{e}
  , \\
  \nabla_{a} {\pounds}_{X} t_{bc}^{\;\;\;\;d}
  &=&
  {\pounds}_{X}\nabla_{a}t_{bc}^{\;\;\;\;d} 
  + X^{e}R_{aeb}^{\;\;\;\;\;\;f} t_{fc}^{\;\;\;\;d} 
  + X^{e}R_{aec}^{\;\;\;\;\;\;f} t_{bf}^{\;\;\;\;d} 
  - X^{e}R_{aef}^{\;\;\;\;\;\;d} t_{bc}^{\;\;\;\;f} 
  \nonumber\\
  && \quad
  + t_{ec}^{\;\;\;\;d}\nabla_{a}\nabla_{b}X^{e} 
  + t_{be}^{\;\;\;\;d}\nabla_{a}\nabla_{c}X^{e} 
  - t_{bc}^{\;\;\;\;e}\nabla_{a}\nabla_{e}X^{d}. 
\end{eqnarray}



\end{document}